\documentclass[prd,
preprint,
superscriptaddress,
nofootinbib,%
tightenlines]{revtex4-2}

\usepackage[%
  colorlinks=true,
  urlcolor=blue,
  linkcolor=blue,
  citecolor=blue
]{hyperref}

\usepackage{slashed,color,amsmath,amssymb,mathrsfs}
\usepackage{graphicx}
\usepackage{subfigure}
\usepackage{booktabs}
\usepackage{hyperref}
\usepackage{longtable}
\usepackage{array}
\usepackage{enumerate}
\usepackage{indentfirst}
\usepackage{verbatim}
\newcommand{\bea}{\begin{eqnarray}} \newcommand{\eea}{\end{eqnarray}}
\allowdisplaybreaks

\begin{document}
\title{Chiral effective Lagrangian for doubly charmed baryons\\ up to $\mathcal{O}(q^4)$}
\author{Peng-Cheng~Qiu}
\affiliation{School of Physics and Electronics, Hunan University, 410082 Changsha, China}
\author{De-Liang Yao}
\email{yaodeliang@hnu.edu.cn}
\affiliation{School of Physics and Electronics, Hunan University, 410082 Changsha, China}

\date{\today}

\begin{abstract}
The chiral effective meson-baryon Lagrangian for the description of interactions between the doubly charmed baryons and Goldstone bosons is constructed up to the order of $q^{4}$. The numbers of linearly independent invariant monomials of $\mathcal{O}(q^2)$, $\mathcal{O}(q^3)$ and $\mathcal{O}(q^4)$ are 8, 32 and 218,  in order.  The obtained Lagrangian can be used to study the  chiral dynamics and relevant phenomenology of the doubly charmed baryons at complete one-loop level in future. For completeness, the non-relativistic reduction of the Lagrangian is also discussed.
\end{abstract}

\maketitle
\newpage
\section{Introduction}
One of the most crucial tasks in hadron physics is to explore hadrons containing heavy quarks, since they are not only necessary for completing the hadron spectroscopy but also useful for our understanding of the QCD dynamics. In recent years, plenty of heavy-flavored baryons have been observed at experiments, some of which are even exotic beyond the expectation of the conventional quark model, e.g. the $P_c$ states as pentaquark candidates reported by LHCb~\cite{Aaij:2015tga,Aaij:2019vzc}.  Though predicted by quark model, members of the spectrum of the doubly and triply heavy-flavored baryons are still absent so far, with the only exception being the $\Xi_{cc}^{++}$ state. The $\Xi_{cc}^{++}$ state is one of the three doubly charmed baryons, showing up in the $\mathbf{20}_M$-plet representation of flavor SU(4) group~\cite{Zyla:2020zbs} concerning $u, d, s, c$ quarks. The three baryons are denoted by $\Xi_{cc}^{++}$, $\Xi_{cc}^+$ and $\Omega_{cc}^{+}$ with quark constituents $[ccu]$, $[ccd]$ and $[ccs]$ in order. 

Nevertheless, the establishment of the existence of the doubly charmed baryons is zigzag. It was first reported in 2002  by SELEX Collaboration~\cite{Mattson:2002vu} that the $\Xi_{cc}^+$ state was observed with measured mass $3519\pm2$~MeV~\cite{Ocherashvili:2004hi}.  However, this baryon state was not confirmed by any other subsequent experimental groups~\cite{Ratti:2003ez,Aubert:2006qw,Chistov:2006zj,Aaij:2013voa}. Moreover, the experimental value is not consistent  with theoretical determinations, e.g., by relativistic quark model~\cite{Ebert:2002ig}, effective potential method~\cite{Karliner:2014gca}, heavy quark effective theory~\cite{Korner:1994nh}, and lattice QCD~\cite{Liu:2009jc,Brown:2014ena}. Actually, the SELEX result is also questionable according to the analysis based on heavy quark-diquark symmetry; see, e.g.,~Ref.~\cite{Brodsky:2011zs}. Hence, the realistic existence of the doubly charmed baryons, especially the $\Xi_{cc}^+$ state, becomes very unclear. The issue was addressed in 2017 that the observation of the doubly charged state $\Xi_{cc}^{++}$ was announced by the LHCb Collaboration~\cite{Aaij:2017ueg}, following the theoretical prediction made by Ref.~\cite{Yu:2017zst}. The reported mass of $\Xi_{cc}^{++}$ is  $3621.4\pm0.78$~MeV,  which is in good agreement with previous theoretical results within 1-$\sigma$ uncertainty~\cite{Karliner:2014gca,Ebert:2002ig,Brown:2014ena}. The finding of the $\Xi_{cc}^{++}$ state has triggered renewed interest in studying doubly charmed baryons, see e.g. Refs.~\cite{Wang:2017azm,Wang:2017mqp,Li:2017pxa,Xiao:2017udy,Li:2017cfz}.  Experiments are still ongoing to investigate the properties of the $\Xi_{cc}^{++}$ state and also to pursue the other two members, i.e. $\Xi_{cc}^{++}$ and $\Omega_{cc}^{+}$, in the family of the doubly charmed baryons. Since the existence of the doubly charmed baryons is now robust, it becomes necessary, in the theoretical side, to investigate them and their excited states using model-independent and systematical methods. 

Chiral perturbation theory (ChPT)~\cite{Weinberg:1978kz,Gasser:1983yg,Gasser:1984gg} is one such method, which plays a prominent role in the study of the low-energy dynamics of QCD, see e.g. Refs.~\cite{Bernard:1995dp,Bernard:2007zu,Epelbaum:2008ga,Yao:2020bxx}.  It is initially developed for the description of the interactions of the Goldstone bosons stemming from the spontaneous breaking of the SU(3)$_L\times$SU(3)$_R$ chiral symmetry of QCD~\cite{Gasser:1984gg}.  The inclusion of  light baryons as degrees of freedom was first done in Ref~\cite{Gasser:1987rb}, and various renormalization versions~\cite{Jenkins:1990jv, Bernard:1992qa,Ellis:1997kc,Becher:1999he,Fuchs:2003qc,Epelbaum:2015vea} have to be proposed to tackle the power counting problem. In order to study heavy-flavored hadron spectrum, ChPT can also be extended to describe the interactions between heavy hadrons and Goldstone bosons. Traditionally,  ChPT for heavy-flavored hadrons was proposed in Refs.~\cite{Burdman:1992gh,Wise:1992hn,Yan:1992gz,Ebert:1994tv,Casalbuoni:1996pg,Hu:2005gf,Colangelo:2012xi}  by implementing heavy-quark symmetry~\cite{Manohar:2000dt} and heavy quark-diquark symmetry~\cite{Hu:2005gf,Meng:2018zbl,Jiang:2019hgs} in addition to chiral symmetry, which means a non-relativistic expansion in terms of the inverse of heavy-flavored hadron mass is performed. However,  such a non-relativistic expansion
distorts the analytic structure of the amplitudes, e.g. the location of the poles of the expanded heavy-flavored hadron propagators are shifted, which could lead to convergence problem. For instance, the scalar form factor of the nucleon at
$t=4 M_\pi^2$ diverges~\cite{Bernard:1996cc,Becher:1999he} . It is thus more appropriate to utilize relativistic treatment of the matter fields involved in ChPT. For doubly charmed baryons, covariant $\chi$PT analyses can be found in Refs.~\cite{Sun:2016wzh,Yao:2018ifh,Blin:2018pmj}, where their masses and electromagnetic form factors were studied at loop level. However, a complete and minimal chiral effective Lagrangian for a full one-loop description of interactions between the doubly charmed baryons and Goldstone bosons is still lacking. In this work, we are going to fill this gap.

This manuscript  is organized as follows. In Sec.~\ref{sec:2},  the relevant chiral building blocks are introduced. In Sec.~\ref{sec:3}, the procedure of the construction of the Lagrangian is described in detail. Transformation properties and chiral dimension of the building blocks together with other necessary ingredients  such as Clifford algebra elements are shown in Sec.~\ref{sec:pro}.  Consequently, invariant monomials are discussed in Sec.~\ref{sec:mono}, while the reduction of the monomials is shown in Sec.~\ref{sec:red}. Our results of the constructed Lagrangian are listed in Sec.~\ref{sec:4},  Appendix~\ref{app:B} and Appendix~\ref{app:C}. A short summary is given in Sec.~\ref{sec:5}.


\section{Chiral building blocks of the Lagrangian}\label{sec:2}
QCD is the underlying theory of ChPT and its Lagrangian reads
\begin{align}
\mathscr{L}=\mathscr{L}^0_{\mathrm{QCD}}+\bar{q}\gamma^{\mu}(\upsilon_{\mu}+\gamma_{5}a_{\mu})q-\bar{q}(s-i\gamma_{5}p)q\ ,
\end{align}
 where $\mathscr{L}^0_{\mathrm{QCD}}$ is the QCD Lagrangian with massless $u$, $d$ and $s$ quarks. The $\mathscr{L}^0_{\mathrm{QCD}}$ exhibits a global $SU(3)_{L} \times SU(3)_{R}$ chiral symmetry, which is spontaneously broken to the subgroup $SU(3)_{V}$ with the emergence of 8 Goldstone bosons according to Goldstone theorem~\cite{Goldstone:1961eq}. Here, $\upsilon_{\mu}$, $a_{\mu}$, $s$, and $p$ are external  vector, axial-vector, scalar and pseudoscalar sources, in order.\footnote{Throughout this work, the vector and axial vector currents should be regarded as traceless $3\times 3$ matrices in the flavor space, i. e. $ \left \langle \upsilon_{\mu} \right \rangle=\left \langle a_{\mu} \right \rangle=0$.}  The underlying Lagrangian possesses a local $SU(3)_{L} \times SU(3)_{R}$ chiral symmetry due to the presence of the external fields. Furthermore, the Goldstone bosons acquire little masses from the explicit breaking of the chiral symmetry  by setting $s={\rm diag}(m_u,m_d,m_s)$.
 
In the chiral effective Lagrangian to be constructed in the following, Goldstone bosons, originating from spontaneously broken chiral symmetry, and the doubly charmed baryons are taken as explicit degrees of freedom. The Goldstone bosons are represented by a matrix $U$, which transforms as 
\bea
U \rightarrow V_{R}UV_{L}^{+}\ ,
\eea
 under chiral transformation, where $V_R$ and $V_L$ are independent SU(3) matrices. The matrix field $U$ is parametrized as 
 \bea
 U=\exp(i\sqrt{2}\Phi/F)\ ,
 \eea
 with $\Phi$ given by
\begin{gather}
\Phi=
\begin{pmatrix} 
\frac{1}{\sqrt{2}}\pi^{0}+\frac{1}{\sqrt{6}}\eta & \pi^{+} & K^{+}\\ \pi^{-}& -\frac{1}{\sqrt{2}}\pi^{0}+\frac{1}{\sqrt{6}}\eta & K^{0} \\
K^{-} & \bar{K}^{0}  & -\frac{2}{\sqrt{6}}\eta
\end{pmatrix},
\end{gather}
where $F$ is the Goldstone-boson decay constant in the $SU(3)$ chiral limit~\cite{Oller:2006yh,book:789407}. The doubly charmed baryons with quantum number $J^{P}=\frac{1}{2}^{+}$ are collected in the triplet
\begin{align}
\psi= \begin{pmatrix}
\Xi^{++}_{cc} \\
 \Xi^{+}_{cc} \\
\Omega^{+}_{cc}
\end{pmatrix},
\end{align}
with $\Xi^{++}_{cc}, \Xi^{+}_{cc}$ and $\Omega^{+}_{cc}$ denoting the doubly charmed baryons. The triplet transforms as
\begin{align}
\psi \rightarrow h(V_{R},V_{L},U)\psi,
\end{align}
where the compensator $h(V_{R},V_{L},U)$ is a nonlinear function of the pion field $U$, $V_{R}$ and $V_{L}$; and it is given by
\begin{align}
h =( \sqrt{V_{R}UV_{L}^{\dagger}} )^{\dagger}V_{R}u\ ,\label{deofh}
\end{align}
with $u=\sqrt{U}$.  It is straightforward to derive the chiral transformation property of the corresponding anti-baryon fields, which reads
\begin{align}
\bar{\psi} \rightarrow \bar{\psi}h^{\dagger}(V_{R},V_{L},U) \ .
\end{align} 
For the construction of the chiral effective Lagrangian,  it is convenient to use building blocks $X$ which transform in a uniform way~\cite{Krause:1990xc}
\begin{align}
X \rightarrow h(V_{R},V_{L},U) X h^{\dagger}(V_{R},V_{L},U)\ ,
\end{align}
The building blocks are linear combinations of pion field and external fields. In our case, the following ones are needed,
\begin{align}
u_{\mu} &=i\{u^{\dagger}(\partial_{\mu}-ir_{\mu})u-u(\partial_{\mu}-il_{\mu})u^{\dagger}\}, \notag\\
f_{\mu\nu}^{\pm} &= uF^{L}_{\mu\nu}u^{\dagger}\pm u^{\dagger}F^{R}_{\mu\nu}u, \notag\\
\chi_{\pm} &= u^{\dagger}\chi u^{\dagger} \pm u\chi^{\dagger} u, 
\end{align}
where
\begin{align}
\chi &=2B_{0}(s+ip),\quad B_{0}=-\left \langle 0|\bar{q}q|0\right \rangle / 3F^{2}, \notag \\
F^{R}_{\mu\nu}&= \partial_{\mu}r_{\nu}-\partial_{\nu}r_{\mu}-i[r_{\nu},r_{\mu}], \quad r_{\mu}= \upsilon_{\mu}+a_{\mu}, \notag \\
F^{L}_{\mu\nu}&= \partial_{\mu}l_{\nu}-\partial_{\nu}l_{\mu}-i[l_{\nu},l_{\mu}], \quad l_{\mu}= \upsilon_{\mu}-a_{\mu}\ .
\end{align}
Here $\left \langle 0|\bar{q}q|0\right \rangle$ denotes the quark condensate,
and $\left \langle \cdots \right \rangle$ stands for trace in the flavour space. 

 The covariant derivative $D^{\mu}$ is defined by
\begin{gather}
D_{\mu} = \partial _{\mu}+\Gamma_{\mu},\notag \\
\Gamma _{\mu}=\frac{1}{2} \{u^{\dagger}(\partial _{\mu}-ir_{\mu})u+u(\partial_{\mu}-il_{\mu})u^{\dagger}  \}.
\end{gather}
For the covariant derivative acting on any building block $X$,  say $[D_{\mu}, X]$, it can be proved that it transforms as 
\bea
[D_{\mu}, X]\rightarrow h[D_{\mu}, X]h^{\dagger}\ ,
\eea
by making use of Eq. \eqref{deofh} and the identity $2(h\Gamma_{\mu}-\Gamma_{\mu}^{\prime}h)=2\partial_{\mu}h
$.\footnote{Here, $\Gamma_{\mu}^{\prime}$ is given by
\begin{align}
\Gamma_{\mu}^{\prime}&=\frac{1}{2} (\sqrt{V_{R}UV_{L}^{\dagger}})^{\dagger}(\partial _{\mu}-iV_{R}r_{\mu}V_{R}^{\dagger}+V_{R}\partial _{\mu}V_{R}^{\dagger})\sqrt{V_{R}UV_{L}^{\dagger}} \notag \\
&+\frac{1}{2} \sqrt{V_{R}UV_{L}^{\dagger}}(\partial _{\mu}-iV_{L}l_{\mu}V_{L}^{\dagger}+V_{L}\partial _{\mu}V_{L}^{\dagger})(\sqrt{V_{R}UV_{L}^{\dagger}})^{\dagger}.
\end{align}
} Nevertheless, for the covariant acting on the baryon field $\psi$, one has
\bea
D_{\mu}\psi \rightarrow h(D_{\mu}\psi)\ ,\quad D_{\mu}\bar{\psi} \rightarrow (D_{\mu}\bar{\psi}) h^{\dagger}\ .
\eea

In addition, the following three relations~\cite{Fettes:1998ud} 
\begin{align}
[D_{\mu},D_{\nu}]X &=\frac{1}{4}[[u_{\mu},u_{\nu}],X]-\frac{i}{2}[f_{\mu\nu}^{+},X],\label{DDX} \\
f_{\mu\nu}^{-}&=[D_{\mu},u_{\nu}]-[D_{\nu},u_{\mu}], \label{DUanti} \\
h_{\mu\nu}&=[D_{\mu},u_{\nu}]+[D_{\nu},u_{\mu}],\label{DU}
\end{align}
are very useful for the construction of the meson-baryon chiral Lagrangian. The element $h_{\mu\nu}$ on the left-hand side of the third relation can  be considered as an extra chiral building block. In consequence, the terms of $[D_{\mu},u_{\nu}]$ and $[D_{\nu},u_{\mu}]$ can be eliminated by using the last two equalities.

\section{Construction of the chiral effective Lagrangian}\label{sec:3}

\subsection{Transformation properties and chiral dimension\label{sec:pro}}
On top of chiral symmetry, the chiral Lagrangian should be invariant under Lorentz transformation,  hermitian conjugation ($h.c.$), discrete $P$ and $C$ symmetries.\footnote{According to CPT theorem,  any Lorentz invariant term one can write down in the Lagrangian is CPT invariant. Hence,  the time reversal invariance is automatically embedded, once Lorentz covariance, hermitian conjugation, spatial inversion and charge conjugation symmetry are implemented in constructing the chiral local Lagrangian.} For easy reference, the transformations properties of the building blocks we use are compiled in Table~\ref{table1}, which are taken from Refs.~\cite{Krause:1990xc, Fettes:1998ud, Gasser:1984gg, Bijnens:1999sh}. Moreover, one has to know the power counting of these elements, listed in the last column of Table~\ref{table1}, such that invariant monomials of the chiral effective Lagrangian can be organized order by order.

\renewcommand{\tablename}{TABLE} 
\renewcommand{\thetable}{\Roman{table}}
\begin{table*}[!h]
\caption{\label{table1}Parity (P), charge conjugation (C), hermitian conjugation ($h.c.$) transformation properties and chiral dimension ($D_\chi$) of the building blocks and the covariant derivative acting on the building blocks. The definitions of $p$, $c$ and $h$ are shown in Eq.~\eqref{dp}, Eq.~\eqref{dc} and Eq.~\eqref{dh}.}
\begin{tabular}{c|ccc|cccc}
\toprule
\midrule
$ $ & $P$ & $C$ & $h.c.$ &$p$&$c$&$h$&$D_\chi$ \\
\bottomrule
$u_{\mu}$ &$ -u^{\mu}$ &$u_{\mu}^{T}$ & $u_{\mu}$ & $1$ & $0$ & $0$ & $1$ \\
$\chi_{+}$ & $\chi_{+}$ & $\chi_{+}^{T}$ & $\chi_{+}$ & $0$ & $0$ & $0$ & $2$ \\
$\chi_{-}$ &$ -\chi_{-}$ &$\chi_{-}^{T}$ & $-\chi_{-}$ & $1$ & $0$ & $1$ & $2$ \\
$f_{+}^{\mu\nu}$ &$ f_{+\mu\nu}$ &$-(f_{+}^{\mu\nu})^{T}$ & $f_{+}^{\mu\nu}$ & $0$ & $1$ & $0$ & $2$ \\
$f_{-}^{\mu\nu}$ &$-f_{-\mu\nu}$ &$(f_{-}^{\mu\nu})^{T}$ & $f_{-}^{\mu\nu}$ & $1$ & $0$ & $0$ & $2$ \\
$h^{\mu\nu}$ &$ -h_{\mu\nu}$ &$(h^{\mu\nu})^{T}$ & $h^{\mu\nu}$ & $1$ & $0$ & $0$ & $2$ \\
$\overrightarrow{D}_{\mu}$ &$\overrightarrow{D}^{\mu}$ &$\overleftarrow{D}_{\mu}^{T}$ & $\overleftarrow{D}_{\mu}$ & $0$ & $0$ & $0$ & $1$ \\
\bottomrule
\bottomrule
\end{tabular}
\end{table*}

Analogously, the transformation properties and power counting of Clifford algebra elements, the imaginary unit, the metric and  Levi-Civita tensors are shown in Table~\ref{table2}, which usually appear as ingredients of baryon bilinear $i\bar{\psi}\Gamma\psi$~\cite{Krause:1990xc}. The covariant derivative acting on the baryon fields is of zeroth chiral order, since the mass of baryons cannot be deemed as a small quantity in the chiral limit.  
\begin{table*}[!h]
\caption{\label{table2}Parity ($P$), charge conjugation ($C$), hermitian conjugation ($h.c.$) transformation properties and chiral dimension ($D_\chi$) of the Clifford algebra elements, the metric and Levi-Civita tensors together with the imaginary unit and the covariant derivative acting on the baryon fields. The definitions of $p$, $c$ and $h$ are shown in Eq.~\eqref{dp}, Eq.~\eqref{dc} and Eq.~\eqref{dh}. }
\begin{tabular}{ccccc}
\toprule
\midrule
$ $  & $p$ &$c$ &$h$ &$D_\chi$ \\
\bottomrule
$i$ & $0$ & $0$ & $1$ & $0$ \\
$1$ & $0$ & $0$ & $0$ & $0$ \\
$\gamma_{5}$ & $1$ & $0$ & $1$ & $1$ \\
$\gamma_{\mu}$ & $0$ & $1$ & $0$ & $0$ \\
$\gamma_{5}\gamma_{\mu}$  & $1$ & $0$ & $0$ & $0$ \\
$\sigma_{\mu\nu}$ & $0$ & $1$ & $0$ & $0$ \\
$g_{\mu\nu}$ & $0$ & $0$ & $0$ & $0$ \\
$\epsilon_{\mu\nu\rho\tau}$  & $1$ & $0$ & $0$ & $0$ \\
$\overrightarrow{D}_{\mu}\psi$ & $0$ & $1$ & $1$ & $0$ \\
\bottomrule
\bottomrule
\end{tabular}
\end{table*}
\subsection{Invariant monomials\label{sec:mono}}
With the elements specified in Table~\ref{table1} and Table~\ref{table2}, we are now in the position to construct all the possible invariant terms. The generic form of any invariant monomial, constrained by Lorentz transformation, chiral symmetry and hermitian conjugation symmetry, in the effective meson-baryon chiral Lagrangian can be written as
\begin{align}\label{generallyform}
(i)^{m}\bar{\psi}A\Gamma D^{n} \psi + h.c., \quad (m=0,1).
\end{align}
Here, 
\begin{enumerate}[(i)]
\item $A$ is a product of building blocks and their covariant derivatives. Since the matrix fields do not commute with each other, one should consider all possible permutations. In addition, it is more preferable to express all the products in terms of combinations of (anti)commutators. For instance, $u_\mu u_\nu$ can be written as $u_\mu u_\nu =(\{u_\mu,u_\nu\}+[u_\mu,u_\nu])/2$.

\item $\Gamma$ is a product of elements of the Clifford algebra basis and/or the metric tensors, the Levi-Civita tensors. More discussions on $\Gamma$ are shown in Appendix~\ref{app:A}.

\item $D^n\equiv\big[D^{\mu_1} D^{\mu_2}\cdots D^{\mu_n}+{\rm all~the~other~permutations}\big]$ is a product of $n\ge 0$ covariant derivatives acting on $\psi$ in a totally symmetrized way.  It is worth pointing out that the commutators of the covariant derivatives can be translated to the basic building blocks with the help of Eq.~\eqref{DDX}.

\end{enumerate}

Note that, in Eq.~\eqref{generallyform}, the Lorentz indices are suppressed for brevity.  However, one needs to keep in mind that all the Lorentz indices of $A$ must be properly contracted with those coming from $\Gamma$ and $D^{n}$ to guarantee  that all the terms are Lorentz scalars. 

Let us proceed with  the transformation property of Eq.~\eqref{generallyform} under parity, which reads
\begin{align}\label{dp}
&\quad \{(i)^{m}\bar{\psi}A\Gamma D^{n} \psi+ h.c.\}^{P} \notag \\
&=(-1)^{p_{A}+p_{\Gamma}+p_{D}}\{(i)^{m}\bar{\psi}A\Gamma D^{n}\psi + h.c.\},
\end{align}
where $p_{A}$ can be determined from Table~\ref{table1},  while $p_{\Gamma}$ and $p_{D}$ from Table~\ref{table2}. It can be concluded from Eq.~\eqref{dp} that the monomial of Eq.~\eqref{generallyform} survives in the chiral effective chiral Lagrangian only if
\begin{align}
(-1)^{p_{A}+p_{\Gamma}+p_{D}}=1.
\end{align}

Although Eq.~\eqref{generallyform} is obviously invariant under hermitic conjugation, it is still worthwhile to demonstrate the corresponding transformation property explicitly. Eq.~\eqref{generallyform} can be rewritten as 
\begin{align}\label{dh}
&\quad (i)^{m}\bar{\psi}A\Gamma D^{n} \psi +(-1)^{h_{A}+h_{\Gamma}+m}(i)^{m}\bar{\psi}\overleftarrow{D}^{n}A\Gamma \psi \notag \\
&=(i)^{m}\bar{\psi}A\Gamma D^{n} \psi +(-1)^{h_{A}+h_{\Gamma}+h_{D}+m}(i)^{m}\bar{\psi}A\Gamma D^{n}\psi + h.o.,
\end{align}
where the values of $h_{A}$, $h_{\Gamma}$ and $h_{D}$ are calculated with the help of Table~\ref{table1} and Table~\ref{table2}. Note that the value of $h_{A}$ includes an additional factor of 1 for each commutator. The terms on the right hand side of Eq.~\eqref{dh} is obtained by making use of integration by parts and the Leibniz rule together with the elimination of total derivatives. The last term $h.o.$ denotes the sum of higher order pieces with covariant derivatives acting on the building blocks shown Table~\ref{table1}, and hence actually can be thrown away from the Lagrangian of a given chiral order under construction. Finally, Eq.~\eqref{generallyform} can occur in the Lagrangian only if
\begin{align}
(-1)^{h_{A}+h_{\Gamma}+h_{D}+m}=1.
\end{align}

Lastly, we check the invariance of Eq.~\eqref{dh}, equivalent to Eq.~\eqref{generallyform}, under charge conjugation,
\begin{align}\label{dc}
&\quad \{2(i)^{m}\bar{\psi}A\Gamma D^{n} \psi +h.o. \}^{C} \notag \\
&=(-1)^{c_{A}+c_{\Gamma}+c_{D}}2(i)^{m}\bar{\psi}A\Gamma D^{n} \psi +h.o.,
\end{align}
where $c_{A}$, which contains an extra factor of 1 for each commutator, is determined from Table~\ref{table1}. Meanwhile, the values of $c_{\Gamma}+c_{D}$ can be obtained from~Table \ref{table2}. Likewise, for a given chiral order,  Eq.~\eqref{generallyform} remains in the Lagrangian only if
\begin{align}
(-1)^{c_{A}+c_{\Gamma}+c_{D}}=1.
\end{align}
\subsection{Reduction of the monomials\label{sec:red}}
A list of invariant monomials, some of them might be linearly dependent, can be obtained according to the procedure discussed above. In this section, we will utilize several linear identities to remove these dependent monomials. The first relation stems from the property of matrix trace, namely
\begin{align}
\left \langle abc  \right \rangle =\left \langle bca \right \rangle =\left \langle cab  \right \rangle.
\end{align}

The second relation is Schouten identity
\begin{align}\label{Schouten}
\epsilon^{\mu\nu\lambda\tau}a^{\rho}+\epsilon^{\nu\lambda\tau\rho}a^{\mu} +\epsilon^{\lambda\tau\rho\mu}a^{\nu} + \epsilon^{\tau\rho\mu\nu}a^{\lambda}+  \epsilon^{\rho\mu\nu\lambda}a^{\tau}=0.
\end{align}

The third relation is obtained by making use of Bianchi identity, Eq.~\eqref{DDX} and Eq.~\eqref{DUanti}, 
\begin{align}
[D_{\mu},f_{\nu\lambda}^{\pm}]+ [D_{\nu},f_{\lambda\mu}^{\pm}]+[D_{\lambda},f_{\mu\nu}^{\pm}]=\frac{i}{2}[u_{\mu},f_{\nu\lambda}^{\mp}] +\frac{i}{2}[u_{\nu},f_{\lambda\mu}^{\mp}] +\frac{i}{2}[u_{\lambda},f_{\mu\nu}^{\mp}].
\end{align}
It can be used to eliminate certain monomials containing $[D_{\mu},f_{\nu\lambda}^{\pm}]$ or $[D_{\nu},f_{\lambda\mu}^{\pm}]$, $[D_{\lambda},f_{\mu\nu}^{\pm}]$, as done in Ref.~\cite{Fettes:2000gb}.

The fourth relation is the so-called Cayley-Hamilton relation~\cite{Bijnens:1999sh}. For any $3\times 3$ arbitrary matrices $a$, $b$ and $c$, the Cayley-Hamilton relation indicates that
\begin{align}
&abc+bca+cab+acb+cba+bac-ab \left \langle c \right \rangle -bc\left \langle a \right \rangle - ca\left \langle b \right \rangle \notag \\
&-ac\left \langle b\right \rangle -cb\left \langle a \right \rangle -ba\left \langle c \right \rangle -a\left \langle bc \right \rangle -b\left \langle ca \right \rangle
-c \left \langle ab \right \rangle -\left \langle abc \right \rangle \notag  \\
&- \left \langle acb \right \rangle +a \left \langle b \right \rangle \left \langle c \right \rangle + b \left \langle c \right \rangle \left \langle a \right \rangle + c\left \langle a \right \rangle \left \langle b \right \rangle + \left \langle a \right \rangle  \left \langle bc \right \rangle + \left \langle b \right \rangle \left \langle ca \right \rangle \notag \\ 
&+ \left \langle  c \right \rangle \left \langle ab \right \rangle - \left \langle a \right \rangle \left \langle b \right \rangle \left \langle c \right \rangle =0\ ,
\end{align}
which is usually adopted to replace the terms with two or more traces by those with one trace or without trace.

On the other hand, it is also possible to use equations  of motion (EOM) to  remove some redundant terms. First of all, the lowest order EOM of the pseudoscalar meson reads 
\begin{align}\label{dumu}
D_{\mu}u^{\mu}=\frac{i}{2} \widetilde{\chi}_{-},
\end{align}
where
\begin{align}
 \widetilde{\chi}_{-}=\chi_{-}-\frac{1}{3}\left \langle \chi_{-} \right \rangle.
\end{align}
With Eq. \eqref{DDX}, Eq.~\eqref{DUanti} and the EOM of Eq. \eqref{dumu}, one can prove that 
 \begin{gather}
D^{2}u_{\mu}=\frac{1}{4}[[u_{\nu},u_{\mu}],u^{\nu}]-\frac{i}{2}[f_{\nu\mu}^{+},u^{\nu}]+D^{\nu}f_{\nu\mu}^{-}+\frac{i}{2}D_{\mu} \widetilde{\chi}_{-}.
\end{gather}
Therefore, $D_{\mu}u^{\mu}$ and $D^{2}u_{\mu}$ can not be regarded as independent structures~\cite{Oller:2006yh, Gasser:1984gg}.

The lowest order EOMs of the baryon fields, obtainable from the Lagrangian in Eq.~\eqref{firstoforder} to be shown in the next section, are
\begin{align}
(i\slashed{D}-m+\frac{g_{A}}{2}\slashed{u}\gamma_{5})\psi=0,\label{eoma}\\
\bar{\psi}(i\overleftarrow{\slashed{D}}+m-\frac{g_{A}}{2}\slashed{u}\gamma_{5}) =0, \label{eomb}
\end{align}
where $(i\slashed{D}-m)$ is counted as $O(q)$. Here $m$ and $g_A$ are the mass and the axial coupling of the doubly charmed baryons in the SU(3) chiral limit, respectively. Based on the above two equations, one can obtain a few linear relations to eliminate many unnecessary terms. The linear relations we use read~\cite{Fettes:1998ud, Fettes:2000gb}:
\begin{align}
\bar{\psi}A^{\mu}iD_{\mu}\psi +h.c.   &\dot{ = }   2m \bar{\psi}\gamma_{\mu}A^{\mu} \psi, \\
\bar{\psi}\gamma_{\mu}[iD^{\mu}, A] \psi &\dot{=} \frac{g_{A}}{2}\bar{\psi}\gamma^{\mu}\gamma_{5}[A,u_{\mu}]\psi,\\
\bar{\psi}\gamma_{5}\gamma_{\mu}[iD^{\mu}, A] \psi &\dot{=} -2m\bar{\psi}\gamma_{5}A\psi -\frac{g_{A}}{2}\bar{\psi}\gamma^{\mu}[A,u_{\mu}]\psi,\\
\bar{\psi}\gamma_{5}\gamma_{\lambda}A^{\mu\lambda}iD_{\mu}\psi +h.c.  &\dot{=} 2im \bar{\psi}\gamma_{5}\sigma_{\mu\lambda}A^{\mu\lambda} \psi \notag \\ 
&+(\bar{\psi}\gamma_{5}\gamma_{\mu}A^{\mu\lambda}iD_{\lambda}\psi +h.c.),\\
\bar{\psi}\gamma_{5}\gamma_{\lambda}A^{\mu\lambda}iD_{\mu}\psi +h.c.  &\dot{=} m \bar{\psi}\sigma^{\nu\rho}\epsilon_{\nu\rho\mu\lambda} A^{\mu\lambda} \psi \notag \\ 
&+(\bar{\psi}\gamma_{5}\gamma_{\mu}A^{\mu\lambda}iD_{\lambda}\psi +h.c.),\\
\bar{\psi}\sigma_{\alpha\beta}A^{\alpha\beta\mu}iD_{\mu}\psi +h.c. & \dot{=} -2m \bar{\psi}\epsilon_{\alpha\beta\mu\nu}\gamma_{5}\gamma^{\nu}A^{\alpha\beta\mu}\psi 
-(\bar{\psi}\sigma_{\beta\mu}A^{\alpha\beta\mu}iD_{\alpha}\psi +h.c.) \notag \\
&+(\bar{\psi}\sigma_{\alpha\mu}A^{\alpha\beta\mu}iD_{\beta}\psi +h.c.), 
\end{align}
where  the symbol $\dot{ = }$ means the two objects on the left- and right-hand sides are equal up to some negligible higher order pieces.
\subsection{Non-relativistic projection }
For completeness, we further consider the doubly charmed baryons as heavy static sources in the non-relativistic limit and perform the so-called heavy baryon (HB) projection~\cite{Jenkins:1990jv} of the relativistic Lagrangian. Here, we present a brief introduction to the non-relativistic approach (for more detailed discussions, see e.g. Refs.~\cite{Jenkins:1990jv, Bernard:1992qa,Bernard:1995dp}).  

The four-momentum of the doubly charmed baryon can be split as
\begin{align}
p_{\mu}=m\upsilon_{\mu} +l_{\mu},
\end{align}
where $\upsilon_{\mu}$ denotes the four-velocity satisfying $\upsilon^{2}=1$ and $l_{\mu}$ is the small off-shell momentum with $ \upsilon \cdot l \ll m $. The doubly charmed baryon field $\psi$ can be decomposed into large component $H$ and small component $h$ via
\begin{align}
\psi =e^{-im\upsilon \cdot x}(H+h),
\end{align}
with
\begin{align}
\slashed{\upsilon}H=H, \quad \slashed{\upsilon}h=-h. 
\end{align}

In terms of $H$ and $h$, the meson-baryon chiral Lagrangian can be recast as 
\begin{align}
\mathscr{L}=\bar{H}\mathcal{A}H+\bar{h}\mathcal{B}H+\bar{H}\gamma_{0}\mathcal{B}^{\dagger}\gamma_{0}h -\bar{h}\mathcal{C}h,
\end{align}
where the operators $\mathcal{A}$, $\mathcal{B}$ and $\mathcal{C}$ may be expanded as a series of the low energy momentum $q$. In this approach, it is more advantageous to make use of the velocity $\upsilon_{\mu}$ and the spin-operator $S_{\mu}=\frac{i}{2}\gamma_{5}\sigma_{\mu\nu}\upsilon^{\nu}$ to express every baryon bilinear $\bar{\psi}\Gamma \psi$. 

After eliminating the small component $h$, the non-relativistic Lagrangian reads
\begin{align}
\mathscr{L}=\bar{H}\{\mathcal{A}+(\gamma_{0}\mathcal{B}^{\dagger}\gamma_{0})\mathcal{C}^{-1}\mathcal{B}\}H,
\end{align}
where
\begin{align}
\mathcal{C}^{-1}=\frac{1}{2m}-\frac{i(\upsilon \cdot D)+g_{A}S \cdot u}{(2m)^{2}}+\frac{(i\upsilon \cdot D+g_{A}S\cdot u)^{2}}{(2m)^{3}}-\frac{\mathcal{C}^{(2)}}{(2m)^{2}}+ \cdots.
\end{align}

\section{The chiral effective meson-baryon Lagrangian}\label{sec:4}
Using the method described above, we have constructed the minimal and complete  chiral effective meson-baryon Lagrangians up to $O(q^{4})$ both in the relativistic and the non-relativistic forms.
\subsection{The Lagrangian at $O(q)$}
The lowest order of relativistic chiral effective Lagrangian reads
\begin{gather}\label{firstoforder}
\mathscr{L}_{M\psi}^{(1)}=\bar{\psi}(i\slashed{D}-m)\psi
+\frac{g_{A}}{2}\bar{\psi}\slashed{u}\gamma_{5}\psi,
\end{gather}
where $m$ is the mass of baryon and $g_{A}$ is the axial-vector coupling constant in the $SU(3)$ chiral limit. The operator $(i\slashed{D}-m)$ is counted as $O(q)$ in the chiral expansion, as discussed in Ref.~\cite{Krause:1990xc}. In principle,  the $g_A$ coupling is an unknown parameter which needs to be determined by experimental data.  In Ref.~\cite{Sun:2016wzh}, it was estimated to be $|g_A|=0.2$. The above leading order Lagrangian has been used to explore the possible exotic states in the spectrum of doubly charmed baryons~\cite{Guo:2017vcf}.

The corresponding non-relativistic Lagrangian can be expressed as
\begin{gather}
\mathscr{\hat{L}}_{M\psi}^{(1)}=\bar{H}(i\upsilon \cdot D +g_{A}S \cdot u)H.
\end{gather}
In the non-relativistic Lagrangian of leading order, the doubly charmed baryon mass term disappears and the Dirac matrices have been substituted by $\upsilon_{\mu}$ and $S_{\mu}$.
\subsection{The Lagrangian at $O(q^{2})$}
The $O(q^{2})$ meson-baryon Lagrangian can be written as
\begin{gather}
\mathscr{L}_{M\psi}^{(2)}=\sum	_{i=1}^{8}b_{i}\bar{\psi}O_{i}^{(2)}\psi ,
\end{gather}
where $b_{i}$'s are unknown low-energy constants (LECs).  It was pointed out by Ref.~\cite{Yan:2018zdt} that some of the $O(q^{2})$ LECs can be related to those in the charmed meson Lagrangian~\cite{Yao:2015qia} by imposing heavy anti-quark-diquark symmetry~\cite{Savage:1990di}.  The monomials $O_{i}^{(2)}$ are given in the 2nd column of  Table~\ref{tab:op2}.\footnote{There is one more term in the $O(q^{2})$ Lagrangian given by Ref.~\cite{Sun:2014aya}. However, terms with $\langle f^+_{\mu\nu} \rangle$ do not show up in our case, due to the fact that the external vector and axial vector currents are set to be traceless.  }

In non-relativistic form, the chiral Lagrangian of $O(q^{2})$ reads
\begin{gather}
\mathscr{\hat{L}}_{M\psi}^{(2)}=\bar{H}\{\mathcal{A}^{(2)}+\gamma_{0}\mathcal{B}^{(1)\dagger}\gamma_{0}\mathcal{C}^{(0)-1}\mathcal{B}^{(1)}\}H,
\end{gather}
where
\begin{align}
\mathcal{A}^{(2)}=\sum_{i=1}^{8}b_{i}\hat{O}_{i}^{(2)}.
\end{align}
The non-relativistic operators $\hat{O}_{i}^{(2)}$ corresponding to the relativistic monomials $O_{i}^{(2)}$ are listed in the 3rd column of Table~\ref{tab:op2}. The second term in the bracket incorporates the $1/m$ corrections, which are shown in Appendix~\ref{app:B}.  Explicit expressions for $\mathcal{B}^{(i)}$ and $\mathcal{C}^{(i)}$ are collected in Appendix~\ref{app:B} as well.  Based on the Lagrangian we obtained here, we have checked that it is straightforward to reproduce the pion-nucleon Lagrangian in Ref.~\cite{Fettes:2000gb} by using the Cayley-Hamilton relation for $2\times 2$ matrices.

\begin{longtable}{ccc}
\caption{Terms in the relativistic and non-relativistic Lagrangian of $O(q^{2})$.} 
\label{tab:op2}\\
\toprule
\midrule
$i$ & $O_{i}^{(2)}$ & $\hat{O}_{i}^{(2)}$ \\
\bottomrule
\endhead
\midrule
\bottomrule
\endfoot
\bottomrule
\bottomrule
\endlastfoot
$1$  & $ \left \langle \chi_{+}\right \rangle$ & $ \left \langle \chi_{+} \right \rangle $\\
$2$  & $\widetilde{\chi}_{+} $ & $\widetilde{\chi}_{+}$ \\
$3$  & $u^{2}$ & $u^{2}$ \\
$4$  & $ \left \langle u^{2}\right \rangle $ & $ \left \langle u^{2} \right \rangle $ \\
$5$  & $\{u^{\mu},u^{\nu}\}D_{\mu\nu} +h.c. $ & $ -8m^{2}(\upsilon \cdot u)^{2}$ \\
$6$  & $\left \langle u^{\mu}u^{\nu} \right \rangle D_{\mu\nu}+h.c. $  & $-4m^{2}\left \langle (\upsilon \cdot u)^{2} \right \rangle $\\
$7$  & $i[u^{\mu},u^{\nu}]\sigma_{\mu\nu} $ & $2[S_{\mu}, S_{\nu}][u^{\mu},u^{\nu}]$\\
$8$  & $f_{+}^{\mu\nu}\sigma_{\mu\nu}$ & $-2i[S_{\mu}, S_{\nu}]f_{+}^{\mu\nu}$ \\
\end{longtable}

\subsection{The Lagrangian at $O(q^{3})$}
The chiral meson-baryon Lagrangian at $O(q^{3})$ takes the form
\begin{gather}
\mathscr{L}_{M\psi}^{(3)}=\sum	_{i=1}^{32}c_{i}\bar{\psi}O_{i}^{(3)}\psi ,
\end{gather}
where $c_{i}$ are $O(q^{3})$ LECs and the operators $O_{i=1,\cdots 32}^{(3)}$ are listed in the 2nd column of Table~\ref{tab:p3}. 

 Also, the $O(q^{3})$ non-relativistic Lagrangian can be written as
\begin{align}
\mathscr{\hat{L}}_{M\psi}^{(3)}=\bar{H}\{\mathcal{A}^{(3)}+\gamma_{0}\mathcal{B}^{(1)\dagger}\gamma_{0}\mathcal{C}^{(0)-1}\mathcal{B}^{(2)}+\gamma_{0}\mathcal{B}^{(1)\dagger}\gamma_{0}\mathcal{C}^{(1)-1}\mathcal{B}^{(1)}+\gamma_{0}\mathcal{B}^{(2)\dagger}\gamma_{0}\mathcal{C}^{(0)-1}\mathcal{B}^{(1)}\}H,
\end{align}
with
\begin{align}
\mathcal{A}^{(3)}=\sum_{i=1}^{32}c_{i}\hat{O}_{i}^{(3)}, 
\end{align}
where the monomials $\hat{O}_{i}^{(3)}$ are collected in the 3rd column of Table~\ref{tab:p3}. The $1/m$ corrections are given in Appendix~\ref{app:B}.
\begin{longtable}{ccc}
\caption{Terms in the $O(q^{3})$ relativistic and non-relativistic Lagrangians.}
\label{tab:p3}\\
\toprule
\midrule
$i$ & $O_{i}^{(3)}$ & $\hat{O_{i}}^{(3)}$ \\
\bottomrule
\endhead
\midrule
\bottomrule
\endfoot
\bottomrule
\bottomrule
\endlastfoot
$1$  & $\{u_{\mu},\{u^{\mu},u^{\nu}\}\}\gamma_{5}\gamma_{\nu}$ & $-2\{u_{\mu},\{u^{\mu},S\cdot u\}\}$\\
$2$  & $[u_{\mu},[u^{\mu},u^{\nu}]]\gamma_{5}\gamma_{\nu}$ &  $-2[u_{\mu},[u^{\mu}, S\cdot u]]$\\
$3$  & $u^{\nu}\gamma_{5}\gamma_{\nu} \left \langle u^{2} \right \rangle  $ & $-2S\cdot u \left \langle u^{2} \right \rangle$ \\
$4$  & $u_{\mu}\gamma_{5}\gamma_{\nu}  \left \langle u^{\mu}u^{\nu} \right \rangle $ & $-2u_{\mu} \left \langle u^{\mu}  S\cdot u \right \rangle $\\
$5$  & $\{u^{\mu},\{u^{\nu},u^{\rho}\}\}\gamma_{5}\gamma_{\mu}D_{\nu\rho} +h.c. $  & $16m^{2}\{S\cdot u, (\upsilon \cdot u)^{2}\}\}$\\
$6$  & $u^{\mu}\gamma_{5}\gamma_{\mu}\left \langle u^{\nu}u^{\rho}\right \rangle D_{\nu\rho}+ h.c. $ & $8m^{2}S\cdot u \left \langle (\upsilon \cdot u)^{2} \right \rangle $ \\
$7$  & $i \epsilon_{\mu\nu\rho\tau}\{[u^{\mu},u^{\nu}],u^{\rho}\} \gamma^{\tau} $ & $i \epsilon_{\mu\nu\rho\tau}\{[u^{\mu},u^{\nu}],u^{\rho}\} \upsilon^{\tau} $ \\
$8$  & $i\epsilon_{\mu\nu\rho\tau}\gamma^{\tau}\left \langle [u^{\mu},u^{\nu}]u^{\rho}\right \rangle $ & $i\epsilon_{\mu\nu\rho\tau}\upsilon^{\tau} \left \langle [u^{\mu},u^{\nu}]u^{\rho}\right \rangle $ \\
$9$  & $i\epsilon_{\mu\nu\lambda\tau}\{u^{\mu},\{u^{\nu},u^{\rho}\}\}\sigma^{\lambda \tau}D_{\rho}+h.c. $  & $-4im\epsilon_{\mu\nu\lambda\tau}\{u^{\mu},\{u^{\nu},\upsilon \cdot u\}\}[S^{\lambda},S^{\tau}] $  \\
$10$  & $i\epsilon_{\mu\nu\lambda\tau}u^{\mu}\sigma^{\lambda \tau} \left \langle u^{\nu}u^{\rho}\right \rangle D_{\rho}+h.c. $ & $-4im\epsilon_{\mu\nu\lambda\tau}u^{\mu} \left \langle u^{\nu}\upsilon \cdot u\right \rangle [S^{\lambda},S^{\tau}]   $ \\
$11$  & $i[u_{\mu},h^{\mu \nu}]\gamma_{\nu} $ & $i[u_{\mu},h^{\mu \nu}]\upsilon_{\nu}$ \\
$12$  & $i[u^{\mu},h^{\nu\rho}]\gamma_{\mu} D_{\nu\rho}+ h.c. $ & $-4im^{2}[\upsilon \cdot u ,h^{\nu\rho}]\upsilon_{\nu} \upsilon_{\rho}$ \\
$13$  & $i\{u^{\mu},h^{\nu\rho}\}\sigma_{\mu\nu}D_{\rho} +h.c. $ & $-4im\{u^{\mu},h^{\nu\rho}\}[S_{\mu}, S_{\nu}] \upsilon_{\rho}$ \\
$14$  & $i\sigma_{\mu\nu} \left \langle u^{\mu}h^{\nu\rho} \right \rangle D_{\rho}+ h.c.$ & $-4im[S_{\mu}, S_{\nu}] \left \langle u^{\mu}h^{\nu\rho} \right \rangle  \upsilon_{\rho}$ \\
$15$  & $\{u^{\mu},\widetilde{\chi}_{+}\}\gamma_{5}\gamma_{\mu}  $ & $-2\{S \cdot u ,\widetilde{\chi}_{+}\}  $ \\
$16$  & $u^{\mu}\gamma_{5}\gamma_{\mu} \left \langle \chi_{+} \right \rangle  $  & $-2S \cdot u \left \langle \chi_{+} \right \rangle  $ \\
$17$  & $\gamma_{5}\gamma_{\mu} \left \langle u^{\mu}\widetilde{\chi}_{+}\right \rangle  $ & $-2 \left \langle S \cdot  u \widetilde{\chi}_{+} \right \rangle  $\\
$18$  & $i\gamma_{5}\gamma_{\mu}[D^{\mu},\widetilde{\chi}_{-}]$ & $-2i[S \cdot D, \widetilde{\chi}_{-}] $\\
$19$  & $i\gamma_{5}\gamma_{\mu}\left \langle [D^{\mu},\chi_{-} ]\right \rangle$ & $-2i\left \langle [S \cdot D, \chi_{-} ]\right \rangle$\\
$20$  & $[\widetilde{\chi}_{-},u^{\mu}]\gamma_{\mu} $ & $[\widetilde{\chi}_{-},\upsilon \cdot u] $ \\
$21$  & $i[u_{\mu},f_{+}^{\mu\nu}]\gamma_{5}\gamma_{\nu}$ & $-2i[u_{\mu},f_{+}^{\mu\nu}]S_{\nu}$\\
$22$  & $\epsilon_{\mu\nu\rho\tau} \{u^{\mu},f_{+}^{\nu\rho}\}\gamma^{\tau} $ & $\epsilon_{\mu\nu\rho\tau} \{u^{\mu},f_{+}^{\nu\rho}\}\upsilon^{\tau}$ \\
$23$  & $\epsilon_{\mu\nu\rho\tau}\gamma^{\tau}  \left \langle u^{\mu}f_{+}^{\nu\rho} \right \rangle $  & $\epsilon_{\mu\nu\rho\tau}\upsilon^{\tau}\left \langle u^{\mu}f_{+}^{\nu\rho} \right \rangle $  \\
$24$  & $\epsilon_{\nu\rho\lambda\tau}[u^{\mu},f^{\nu \rho}_{+}]\sigma^{\lambda \tau}D_{\mu} +h.c.$ & $-4m\epsilon_{\nu\rho\lambda\tau}[\upsilon \cdot u,f^{\nu \rho}_{+}][S^{\lambda},S^{\tau}]$\\
$25$  & $i[D_{\mu},f^{\mu\nu}_{+}]D_{\nu}+h.c.$ & $2m[D_{\mu},f^{\mu\nu}_{+}]\upsilon_{\nu}$  \\
$26$  & $i[u_{\mu},f^{\mu \nu}_{-}]\gamma_{\nu}$ & $i[u_{\mu},f^{\mu \nu}_{-}]\upsilon_{\nu} $ \\
$27$  & $\epsilon_{\mu\nu\rho\tau} \{u^{\mu},f_{-}^{\nu\rho}\}\gamma_{5}\gamma^{\tau}$ & $-2\epsilon_{\mu\nu\rho\tau} \{u^{\mu},f_{-}^{\nu\rho}\}S^{\tau} $ \\
$28$  & $\epsilon_{\mu\nu\rho\tau}\gamma_{5}\gamma^{\tau} \left \langle u^{\mu}f_{-}^{\nu\rho} \right  \rangle   $  & $-2\epsilon_{\mu\nu\rho\tau}S^{\tau} \left \langle u^{\mu}f_{-}^{\nu\rho} \right  \rangle   $ \\
$29$  & $i\{u^{\mu},f^{\nu\rho}_{-}\}\sigma_{\mu\nu}D_{\rho}+ h.c. $ & $-4im\{u^{\mu},f^{\nu\rho}_{-}\}[S_{\mu},S_{\nu}]\upsilon_{\rho}$\\
$30$  & $i\sigma_{\mu\nu} \left \langle u^{\mu}f^{\nu\rho}_{-}\right  \rangle D_{\rho}+ h.c. $ & $-4im[S_{\mu},S_{\nu}]\upsilon_{\rho} \left \langle u^{\mu}f^{\nu\rho}_{-}\right  \rangle  $\\
$31$  & $[D_{\mu},f^{\mu\nu}_{-}]\gamma_{5}\gamma_{\nu}$ & $-2S_{\nu}[D_{\mu},f^{\mu\nu}_{-}]$\\
$32$  & $[D^{\lambda},f^{\mu\nu}_{-}]\gamma_{5}\gamma_{\mu}D_{\nu\lambda}+ h.c.$ & $8m^{2}S_{\mu}[\upsilon \cdot D,f^{\mu\nu}_{-}]\upsilon_{\nu}$ \\
\end{longtable}

\subsection{The Lagrangian at $O(q^{4})$}
The chiral effective meson-baryon Lagrangian at $O(q^{4})$ reads
\begin{gather}
\mathscr{L}_{M\psi}^{(4)}=\sum	_{i=1}^{218}d_{i}\bar{\psi}O_{i}^{(4)}\psi,
\end{gather}
where $d_{i}$ are LECs and the terms of $O(q^{4})$ are tabulated in the 2nd column of Table~\ref{tab:p4}.  The last three terms are contact terms (See~Refs.~\cite{Bijnens:1999sh, Jiang:2016vax} for analogous operators). {It should be noted that they are combinations depending only on external fields and therefore are not directly accessible experimentally. Furthermore,  as pointed out by Ref.~\cite{Bijnens:1999sh}, it is more convenient to express the contact terms in terms of the LR-basis, i.e. $F^{\mu\nu}_{R/L}$, and the $\chi$ element, rather than those chiral building blocks in Table~\ref{table1}.
}

The $O(q^{4})$ meson-baryon non-relativistic Lagrangian which contains the $1/m$ corrections takes the form
\begin{align}
\mathscr{\hat{L}}_{M\psi}^{(4)}&=\bar{H}\{\mathcal{A}^{(4)}+\gamma_{0}\mathcal{B}^{(1)\dagger}\gamma_{0}\mathcal{C}^{(0)-1}\mathcal{B}^{(3)}+\gamma_{0}\mathcal{B}^{(1)\dagger}\gamma_{0}\mathcal{C}^{(1)-1}\mathcal{B}^{(2)}+\gamma_{0}\mathcal{B}^{(1)\dagger}\gamma_{0}\mathcal{C}^{(2)-1}\mathcal{B}^{(1)} \notag \\
&+\gamma_{0}\mathcal{B}^{(2)\dagger}\gamma_{0}\mathcal{C}^{(0)-1}\mathcal{B}^{(2)} +\gamma_{0}\mathcal{B}^{(2)\dagger}\gamma_{0}\mathcal{C}^{(1)-1}\mathcal{B}^{(1)}+\gamma_{0}\mathcal{B}^{(3)\dagger}\gamma_{0}\mathcal{C}^{(0)-1}\mathcal{B}^{(1)}\}H,
\end{align}
with
\begin{align}
\mathcal{A}^{(3)}=\sum_{i=1}^{218}d_{i}\hat{O}_{i}^{(4)},
\end{align}
where the operators $\hat{O}_{i}^{(4)}$ are compiled in the 3rd column of Table~\ref{tab:p4}.

\section{Summary}\label{sec:5}
Based on chiral symmetry and basic invariances such as Lorentz invariance,  we have constructed the chiral effective Lagrangian for the description of the interactions between the doubly charmed ground-state baryons and Goldstone bosons up to $O(q^{4})$.  Complete and minimal sets of $O(q^{3})$ and $O(q^{4})$ operators are established for the first time. The numbers of $O(q^{3})$ and $O(q^{4})$ terms are 32 and 218, respectively. The involved LECs are expected to be determined by, e.g. experimental or lattice QCD data in future. The obtained Lagrangian is sufficient for comprehensive analyses of the low-energy physics of the doubly charmed baryons up to the fourth order,  enabling us to explore the doubly charmed spectroscopy with high accuracy. Furthermore, it can be readily extended to the sector of doubly bottomed baryons according to heavy quark flavor symmetry.

\section*{Acknowledgments}
We would like to thank Qin-He Yang for useful discussions. This work is supported by National Nature Science Foundations of China (NSFC) under Contract No. 11905258 and by the Fundamental Research Funds for the Central Universities.
\appendix
\section{\label{app:A}Types of $\Theta_{\mu\nu \dots}$}
In this Appendix,  a brief introduction to the EOM constraints on the chiral Lagrangian is presented; see Refs.~\cite{Fettes:2000gb, Jiang:2016vax} for more discussions. $\Gamma$ can be one of the Clifford algebra elements \{$1$, $\gamma_{5}\gamma_{\mu}$, $\sigma_{\mu\nu}$\} or the Levi-Civita tensor $\epsilon_{\mu\nu\lambda\tau}$.
\begin{itemize}
\item For the Clifford algebra elements \{$1$, $\gamma_{5}\gamma_{\mu}$, $\sigma_{\mu\nu}$\}, their indices should be distinguished from those of the covariant derivatives that act on the baryon field $\psi$.
\item For $\epsilon_{\mu\nu\lambda\tau}$, only one of its indices can be contracted with those of the covariant derivatives that act on the baryon field $\psi$.
\end{itemize}

Therefore, it is more convenient to use $\Theta_{\mu\nu \dots} =(\Gamma D^{n})_{\mu\nu\dots}$, taken from Ref.~\cite{Jiang:2016vax}, to construct our chiral Lagrangian. The types of $\Theta_{\mu\nu \dots} $ we need are as follows:
\begin{align}
&1;\notag\\
&\gamma_{5}\gamma_{\mu}, D_{\mu};\notag\\
&\sigma_{\mu\nu}, \gamma_{5}\gamma_{\mu}D_{\nu}, D_{\mu\nu}; \notag\\
&\gamma_{5}\gamma_{\mu}D_{\nu\lambda}, \sigma_{\mu\nu}D_{\lambda}, \epsilon_{\mu\nu\lambda\rho}D^{\rho}, D_{\mu\nu\lambda}; \notag\\
&\epsilon_{\mu\nu\lambda\rho}, \gamma_{5}\gamma_{\mu}D_{\nu\lambda\rho}, \sigma_{\mu\nu}D_{\lambda\rho}, \epsilon_{\mu\nu\lambda\tau}{D^{\tau}}_{\rho}, D_{\mu\nu\lambda\rho}.
\end{align}

\section{\label{app:B}Some explicit expressions for non-relativistic Lagrangian }
Basic relations regarding the covariant spin-operator $S_{\mu}$ are given below ($\epsilon^{0123}=-1$):
\begin{align}
S \cdot \upsilon =0,\quad  \{S_{\mu},S_{\nu}\}= \frac{1}{2}(\upsilon_{\mu}\upsilon_{\nu}-g_{\mu\nu}),\quad [S_{\mu},S_{\nu}]=i\epsilon_{\mu\nu\alpha\beta}\upsilon^{\alpha}S^{\beta}.
\end{align}

By using the above equalities, the Dirac bi-linears can be rewritten as 
\begin{align}
&\bar{H}\gamma_{\mu}H=\upsilon_{\mu}\bar{H}H, &  &\bar{h}\gamma_{\mu}H=-2\bar{h}\gamma_{5}S_{\mu}H, \notag \\
&\bar{H}\gamma_{\mu}\gamma_{5}H=2\bar{H}S_{\mu}H, &  &\bar{h}\gamma_{\mu}\gamma_{5}H=-\upsilon_{\mu}\bar{h}\gamma_{5}H, \notag \\
&\bar{H}\sigma^{\mu\nu}H=-2i\bar{H}[S^{\mu},S^{\nu}]H, & &\bar{h}\sigma^{\mu\nu}H=2i\bar{h}\gamma_{5}(\upsilon^{\mu}S^{\nu}-\upsilon^{\nu}S^{\mu})H. 
\end{align}
Subsequently, we can readily obtain some relevant explicit expressions of $\mathcal{B}^{(i)}$ and $\mathcal{C}^{(i)}$:
\begin{align}
\mathcal{B}^{(1)}&=-2i\gamma_{5} S \cdot D -\frac{g_{A}}{2}\gamma_{5} \upsilon \cdot u,\\
\mathcal{B}^{(2)}&=-2b_{7}\gamma_{5}(\upsilon^{\mu}S^{\nu}-\upsilon^{\nu}S^{\mu})[u_{\mu},u_{\nu}]+2ib_{8}\gamma_{5}(\upsilon^{\mu}S^{\nu}-\upsilon^{\nu}S^{\mu})f_{+\mu\nu},\\
\mathcal{B}^{(3)}&=\gamma_{5}(c_{1}\{u_{\mu},\{u^{\mu},\upsilon \cdot u\}\}+c_{2}[u_{\mu},[u^{\mu}, \upsilon \cdot u]]+c_{3}\upsilon \cdot u \left \langle u^{2} \right \rangle +c_{4}u_{\mu} \left \langle u^{\mu}  \upsilon \cdot u \right \rangle  \notag \\ 
&-16m^{2}c_{5}(\upsilon \cdot u)^{3}-4m^{2}c_{6}\upsilon \cdot u \left \langle (\upsilon \cdot u)^{2} \right \rangle +c_{15}\{\upsilon \cdot u ,\widetilde{\chi}_{+}\} +c_{16}\upsilon \cdot u \left \langle \chi_{+} \right \rangle \notag \\
&+c_{17}\left \langle \upsilon \cdot  u \widetilde{\chi}_{+} \right \rangle +ic_{18}[\upsilon \cdot D, \widetilde{\chi}_{-}] +ic_{19}\left \langle [\upsilon \cdot D, \chi_{-}] \right \rangle +ic_{21}[u_{\mu},f_{+}^{\mu\nu}]\upsilon_{\nu} \notag \\
 &+c_{27}\epsilon_{\mu\nu\rho\tau} \{u^{\mu},f_{-}^{\nu\rho}\}\upsilon^{\tau}+c_{28}\epsilon_{\mu\nu\rho\tau}\upsilon^{\tau} \left \langle u^{\mu}f_{-}^{\nu\rho} \right  \rangle +c_{31}\upsilon_{\nu}[D_{\mu},f^{\mu\nu}_{-}] \notag \\
 &-4m^{2}c_{32}[\upsilon \cdot D,f^{\mu\nu}_{-}]\upsilon_{\mu}\upsilon_{\nu} )-2i\epsilon_{\mu\nu\rho\tau}\gamma_{5}S^{\tau}(c_{7}\{[u^{\mu},u^{\nu}],u^{\rho}\}+c_{8}\left \langle [u^{\mu},u^{\nu}]u^{\rho}\right \rangle)\notag \\
 &+4im\epsilon_{\mu\nu\lambda\tau} \gamma_{5}(\upsilon^{\lambda}S^{\tau}-\upsilon^{\tau}S^{\lambda})\{c_{9}\{u^{\mu},\{u^{\nu},\upsilon \cdot u\}\}+c_{10}u^{\mu}\left \langle u^{\nu}\upsilon \cdot u\right \rangle \} \notag \\
&-2ic_{11}[u_{\mu},h^{\mu \nu}]\gamma_{5}S_{\nu} +8im^{2}c_{12}\gamma_{5}[S \cdot u ,h^{\nu\rho}]\upsilon_{\nu} \upsilon_{\rho}+4im\gamma_{5}(\upsilon_{\mu}S_{\nu}-\upsilon_{\nu}S_{\mu})\upsilon_{\rho}\notag \\
&(c_{13}\{u^{\mu},h^{\nu\rho}\}+c_{14}\left \langle u^{\mu}h^{\nu\rho} \right \rangle +c_{29}\{u^{\mu},f^{\nu\rho}_{-}\}+c_{30} \left \langle u^{\mu}f^{\nu\rho}_{-}\right  \rangle)-2c_{20}\gamma_{5}[\widetilde{\chi}_{-}, S \cdot u] \notag \\
&-2\epsilon_{\mu\nu\rho\tau}\gamma_{5}S^{\tau}(c_{22} \{u^{\mu},f_{+}^{\nu\rho}\}+c_{23}\left \langle u^{\mu}f_{+}^{\nu\rho} \right \rangle)+4mc_{24}\epsilon_{\nu\rho\lambda\tau}[\upsilon \cdot u,f^{\nu \rho}_{+}]\gamma_{5}(\upsilon^{\lambda}S^{\tau}-\upsilon^{\tau}S^{\lambda}) \notag \\
&-2ic_{26}[u_{\mu},f^{\mu \nu}_{-}]\gamma_{5}S_{\nu}.
\end{align}
\begin{align}
\mathcal{C}^{(0)}&=2m,\\
\mathcal{C}^{(1)}&=i\upsilon \cdot D +g_{A}S \cdot u,\\
\mathcal{C}^{(2)}&=-b_{1}\left \langle \chi_{+} \right \rangle -b_{2} \widetilde{\chi}_{+}-b_{3}u^{2}-b_{4}\left \langle u^{2} \right \rangle +8m^{2}b_{5}(\upsilon \cdot u)^{2}+4m^{2}b_{6}\left \langle (\upsilon \cdot u)^{2} \right \rangle \notag \\
&-2b_{7}[S^{\mu}, S^{\nu}][u_{\mu},u_{\nu}]+2ib_{8}[S_{\mu}, S_{\nu}]f_{+}^{\mu\nu}.
\end{align}
\begin{align}
\gamma_{0}\mathcal{B}^{(1)\dagger}\gamma_{0}\mathcal{C}^{(0)-1}\mathcal{B}^{(1)}&= \frac{2}{m}(S \cdot D)^{2}-\frac{ig_{A}}{2m}\{S\cdot D,\upsilon \cdot u\} -\frac{g_{A}^{2}}{8m}(\upsilon \cdot u)^{2} ,\\
\gamma_{0}\mathcal{B}^{(1)\dagger}\gamma_{0}\mathcal{C}^{(0)-1}\mathcal{B}^{(2)}&=-\frac{2i}{m}b_{7}S \cdot D (\upsilon^{\mu}S^{\nu}-\upsilon^{\nu}S^{\mu})[u_{\mu},u_{\nu}] -\frac{2}{m}b_{8}S \cdot D (\upsilon^{\mu}S^{\nu}-\upsilon^{\nu}S^{\mu})f_{+\mu\nu}\notag \\
&+\frac{ig_{A}}{2m}b_{8}(\upsilon^{\mu}S^{\nu}-\upsilon^{\nu}S^{\mu})\upsilon \cdot u f_{+\mu\nu}-\frac{g_{A}}{2m}b_{7}(\upsilon^{\mu}S^{\nu}-\upsilon^{\nu}S^{\mu})\upsilon \cdot u [u_{\mu},u_{\nu}] ,\\
\gamma_{0}\mathcal{B}^{(1)\dagger}\gamma_{0}\mathcal{C}^{(1)-1}\mathcal{B}^{(1)}&=-\frac{i}{m^{2}}S \cdot D \upsilon \cdot D S \cdot D -\frac{g_{A}}{4m^{2}} \upsilon \cdot u  \upsilon \cdot D S \cdot D -\frac{g_{A}}{4m^{2}} S \cdot D \upsilon \cdot D  \upsilon \cdot u \notag \\
&+\frac{ig_{A}^{2}}{16m^{2}} \upsilon \cdot u \upsilon \cdot D \upsilon \cdot u-\frac{g_{A}}{m^{2}}S \cdot D S \cdot u S \cdot D +\frac{ig_{A}^{2}}{4m^{2}}  \upsilon \cdot u S \cdot u S \cdot D \notag \\
&+ \frac{ig_{A}^{2}}{4m^{2}} S \cdot D  S \cdot u  \upsilon \cdot u +\frac{g_{A}^{3}}{16m^{2}}\upsilon \cdot u S \cdot u \upsilon \cdot u ,\\
\gamma_{0}\mathcal{B}^{(2)\dagger}\gamma_{0}\mathcal{C}^{(0)-1}\mathcal{B}^{(1)}&=\frac{2i}{m}b_{7}(\upsilon^{\mu}S^{\nu}-\upsilon^{\nu}S^{\mu})[u_{\mu},u_{\nu}] S \cdot D +\frac{2}{m}b_{8}(\upsilon^{\mu}S^{\nu}-\upsilon^{\nu}S^{\mu})f_{+\mu\nu} S \cdot D \notag \\
&-\frac{ig_{A}}{2m}b_{8}(\upsilon^{\mu}S^{\nu}-\upsilon^{\nu}S^{\mu}) f_{+\mu\nu} \upsilon \cdot u+\frac{g_{A}}{2m}b_{7}(\upsilon^{\mu}S^{\nu}-\upsilon^{\nu}S^{\mu})[u_{\mu},u_{\nu}] \upsilon \cdot u  ,\\
\gamma_{0}\mathcal{B}^{(1)\dagger}\gamma_{0}\mathcal{C}^{(0)-1}\mathcal{B}^{(3)}&=-4 \epsilon_{\mu\nu\lambda\tau}S \cdot D(\upsilon^{\lambda}S^{\tau}-\upsilon^{\tau}S^{\lambda})(c_{9}\{u^{\mu},\{u^{\nu},\upsilon \cdot u\}\}+c_{10}u^{\mu}\left \langle u^{\nu}\upsilon \cdot u\right \rangle )\notag \\
&-4S \cdot D(\upsilon_{\mu}S_{\nu}-\upsilon_{\nu}S_{\mu})\upsilon_{\rho}(c_{13}\{u^{\mu},h^{\nu\rho}\} +c_{14}\left \langle u^{\mu}h^{\nu\rho} \right \rangle +c_{29}\{u^{\mu},f^{\nu\rho}_{-}\} \notag \\
&+c_{30} \left \langle u^{\mu}f^{\nu\rho}_{-}\right  \rangle  ) +4ic_{24}\epsilon_{\nu\rho\lambda\tau}S \cdot D[\upsilon \cdot u,f^{\nu \rho}_{+}](\upsilon^{\lambda}S^{\tau}-\upsilon^{\tau}S^{\lambda})  \notag \\
&+\frac{1}{m}S \cdot D(ic_{1}\{u_{\mu},\{u^{\mu},\upsilon \cdot u\}\}  +ic_{2}[u_{\mu},[u^{\mu}, \upsilon \cdot u]] +2c_{11}[u_{\mu},h^{\mu \nu}]S_{\nu}\notag \\
&+ic_{15}\{\upsilon \cdot u ,\widetilde{\chi}_{+}\}+ic_{16}\upsilon \cdot u \left \langle \chi_{+} \right \rangle +ic_{17}\left \langle \upsilon \cdot  u \widetilde{\chi}_{+} \right \rangle -c_{18}[\upsilon \cdot D, \widetilde{\chi}_{-}] \notag \\
&-c_{19}\left \langle [\upsilon \cdot D, \chi_{-}] \right \rangle -2ic_{20}[\widetilde{\chi}_{-}, S \cdot u] -c_{21}[u_{\mu},f_{+}^{\mu\nu}]\upsilon_{\nu}  +2c_{26}[u_{\mu},f^{\mu \nu}_{-}]S_{\nu} )\notag \\
&-\frac{2i}{m}\epsilon_{\mu\nu\rho\tau} S \cdot D S^{\tau}(c_{22}\{u^{\mu},f_{+}^{\nu\rho}\}+c_{23}\left \langle u^{\mu}f_{+}^{\nu\rho} \right \rangle)+\frac{i}{m}\epsilon_{\mu\nu\rho\tau} S \cdot D   \notag \\
&(c_{27}\{u^{\mu},f_{-}^{\nu\rho}\}\upsilon^{\tau}+c_{28} \upsilon^{\tau} \left \langle u^{\mu}f_{-}^{\nu\rho} \right  \rangle)+\frac{i}{m}S \cdot D (c_{3} \upsilon \cdot u \left \langle u^{2} \right \rangle \notag \\
&+c_{31} \upsilon_{\nu}[D_{\mu},f^{\mu\nu}_{-}] +c_{4}u_{\mu} \left \langle u^{\mu}  \upsilon \cdot u \right \rangle) +\frac{2}{m}c_{7}\epsilon_{\mu\nu\rho\tau}S \cdot D\{[u^{\mu},u^{\nu}],u^{\rho}\} S^{\tau}  \notag \\
&+\frac{2}{m}c_{8}\epsilon_{\mu\nu\rho\tau}S \cdot D S^{\tau} \left \langle [u^{\mu},u^{\nu}]u^{\rho}\right \rangle -8mc_{12}S \cdot D[S \cdot u ,h^{\nu\rho}]\upsilon_{\nu} \upsilon_{\rho}  \notag \\
&-4imc_{32}S \cdot D [\upsilon \cdot D,f^{\mu\nu}_{-}]\upsilon_{\mu}\upsilon_{\nu}-4imc_{5}S \cdot D \{4c_{5}(\upsilon \cdot u)^{3} \notag \\
&+c_{6}\upsilon \cdot u \left \langle (\upsilon \cdot u)^{2} \right \rangle \}+ig_{A}\epsilon_{\mu\nu\lambda\tau}(\upsilon^{\lambda}S^{\tau}-\upsilon^{\tau}S^{\lambda})(c_{9}\upsilon \cdot u\{u^{\mu},\{u^{\nu},\upsilon \cdot u\}\}\notag \\
&+c_{10}\upsilon \cdot u u^{\mu}\left \langle u^{\nu}\upsilon \cdot u\right \rangle)+ig_{A}c_{13}\upsilon \cdot u (\upsilon_{\mu}S_{\nu}-\upsilon_{\nu}S_{\mu})\{u^{\mu},h^{\nu\rho}\}\upsilon_{\rho} \notag \\
&+ig_{A}c_{14}\upsilon \cdot u (\upsilon_{\mu}S_{\nu}-\upsilon_{\nu}S_{\mu})\left \langle u^{\mu}h^{\nu\rho} \right \rangle  \upsilon_{\rho} +g_{A}c_{24}\epsilon_{\nu\rho\lambda\tau}\upsilon \cdot u [\upsilon \cdot u,f^{\nu \rho}_{+}] \notag \\
&(\upsilon^{\lambda}S^{\tau}-\upsilon^{\tau}S^{\lambda})+ig_{A}c_{29}\upsilon \cdot u (\upsilon_{\mu}S_{\nu}-\upsilon_{\mu}S_{\nu})\upsilon_{\rho} \{u^{\mu},f^{\nu\rho}_{-}\} \notag \\
&+ig_{A}c_{30}\upsilon \cdot u (\upsilon_{\mu}S_{\nu}-\upsilon_{\mu}S_{\nu})\upsilon_{\rho} \left \langle u^{\mu}f^{\nu\rho}_{-}\right  \rangle +\frac{g_{A}}{4m}c_{1}\upsilon \cdot u \{u_{\mu},\{u^{\mu},\upsilon \cdot u\}\}   \notag \\
&-\frac{ig_{A}}{2m}c_{11}\upsilon \cdot u [u_{\mu},h^{\mu \nu}]S_{\nu} +\frac{g_{A}}{4m}\upsilon \cdot u (c_{15}\{\upsilon \cdot u ,\widetilde{\chi}_{+}\} +c_{16}\upsilon \cdot u \left \langle \chi_{+} \right \rangle \notag \\
&+c_{17}\left \langle \upsilon \cdot  u \widetilde{\chi}_{+} \right \rangle +ic_{18} [\upsilon \cdot D, \widetilde{\chi}_{-}] )+
\frac{g_{A}}{4m}c_{2}\upsilon \cdot u [u_{\mu},[u^{\mu}, \upsilon \cdot u]] \notag \\
&+\frac{ig_{A}}{4m}c_{19}\upsilon \cdot u \left \langle [\upsilon \cdot D, \chi_{-}] \right \rangle -\frac{g_{A}}{2m}c_{20}\upsilon \cdot u [\widetilde{\chi}_{-}, S \cdot u]   \notag \\
&+\frac{ig_{A}}{4m}c_{21}\upsilon \cdot u [u_{\mu},f_{+}^{\mu\nu}]\upsilon_{\nu} -\frac{g_{A}}{2m}\epsilon_{\mu\nu\rho\tau}\upsilon \cdot u S^{\tau}(c_{22}\{u^{\mu},f_{+}^{\nu\rho}\}+c_{23}\left \langle u^{\mu}f_{+}^{\nu\rho} \right \rangle) \notag \\
&-\frac{ig_{A}}{2m}c_{26}\upsilon \cdot u [u_{\mu},f^{\mu \nu}_{-}]S_{\nu} +\frac{g_{A}}{4m}c_{27}\epsilon_{\mu\nu\rho\tau} \upsilon \cdot u \{u^{\mu},f_{-}^{\nu\rho}\}\upsilon^{\tau} \notag \\ 
&+\frac{g_{A}}{4m}c_{28}\epsilon_{\mu\nu\rho\tau}\upsilon \cdot u \upsilon^{\tau} \left \langle u^{\mu}f_{-}^{\nu\rho} \right  \rangle +\frac{g_{A}}{4m}c_{3}(\upsilon \cdot u)^{2}  \left \langle u^{2} \right \rangle \notag \\
&+\frac{g_{A}}{4m}c_{31}\upsilon \cdot u \upsilon_{\nu}[D_{\mu},f^{\mu\nu}_{-}]  +\frac{g_{A}}{4m}c_{4}\upsilon \cdot u u_{\mu} \left \langle u^{\mu}  \upsilon \cdot u \right \rangle  \notag \\
&-\frac{ig_{A}}{2m}c_{7}\epsilon_{\mu\nu\rho\tau}\upsilon \cdot u \{[u^{\mu},u^{\nu}],u^{\rho}\} S^{\tau}-\frac{ig_{A}}{2m}c_{8}\epsilon_{\mu\nu\rho\tau}\upsilon \cdot u S^{\tau} \left \langle [u^{\mu},u^{\nu}]u^{\rho}\right \rangle  \notag \\
&+2img_{A}c_{12}\upsilon \cdot u [S \cdot u ,h^{\nu\rho}]\upsilon_{\nu} \upsilon_{\rho}  -mg_{A}c_{32}\upsilon \cdot u [\upsilon \cdot D,f^{\mu\nu}_{-}]\upsilon_{\mu}\upsilon_{\nu}\notag \\
&-4mg_{A}c_{5}(\upsilon \cdot u)^{4}  -mg_{A}c_{6}(\upsilon \cdot u)^{2}  \left \langle (\upsilon \cdot u)^{2} \right \rangle , \\
\gamma_{0}\mathcal{B}^{(1)\dagger}\gamma_{0}\mathcal{C}^{(1)-1}\mathcal{B}^{(2)}&=-\frac{1}{m^{2}}b_{7}S \cdot D  (\upsilon \cdot D-ig_{A}S \cdot u ) (\upsilon^{\mu}S^{\nu}-\upsilon^{\nu}S^{\mu})[u_{\mu},u_{\nu}] \notag \\
&+\frac{ig_{A}}{4m^{2}}b_{7}\upsilon \cdot u (\upsilon \cdot D -ig_{A}S \cdot u)(\upsilon^{\mu}S^{\nu}-\upsilon^{\nu}S^{\mu})[u_{\mu},u_{\nu}] \notag \\
&+ \frac{i}{m^{2}}b_{8}S \cdot D (\upsilon \cdot D-ig_{A}S \cdot u) (\upsilon^{\mu}S^{\nu}-\upsilon^{\nu}S^{\mu})f_{+\mu\nu}   \notag \\
&+\frac{g_{A}}{4m^{2}}b_{8}\upsilon \cdot u( \upsilon \cdot D-ig_{A}S \cdot u) (\upsilon^{\mu}S^{\nu}-\upsilon^{\nu}S^{\mu})f_{+\mu\nu} , \\
\gamma_{0}\mathcal{B}^{(1)\dagger}\gamma_{0}\mathcal{C}^{(2)-1}\mathcal{B}^{(1)}&=-\frac{1}{2m^{3}}S \cdot D (\upsilon \cdot D)^{2} S \cdot D+\frac{ig_{A}}{2m^{3}}S \cdot D \{\upsilon \cdot D, S \cdot u \}S \cdot D\notag \\
&+\frac{g_{A}^{2}}{2m^{3}}S \cdot D(S \cdot u )^{2} S \cdot D-\frac{1}{m^{2}}S \cdot D C_{(2)} S \cdot D\notag \\
&+\frac{g_{A}^{2}}{32m^{3}}\upsilon \cdot u (\upsilon \cdot D)^{2} \upsilon \cdot u -\frac{ig_{A}^{3}}{32m^{3}}\upsilon \cdot u \{\upsilon \cdot D, S \cdot u \} \upsilon \cdot u \notag \\
&-\frac{g_{A}^{4}}{32m^{3}}\upsilon \cdot u (S \cdot u )^{2} \upsilon \cdot u +\frac{g_{A}^{2}}{16m^{2}}\upsilon \cdot u C_{(2)} \upsilon \cdot u \notag \\
&+\frac{ig_{A}}{8m^{3}}S \cdot D (\upsilon \cdot D)^{2} \upsilon \cdot u +\frac{g_{A}^{2}}{8m^{3}}S \cdot D \{\upsilon \cdot D, S \cdot u \} \upsilon \cdot u \notag \\
&-\frac{ig_{A}^{3}}{8m^{3}}S \cdot D (S \cdot u )^{2} \upsilon \cdot u+\frac{ig_{A}}{4m^{2}}S \cdot D C_{(2)} \upsilon \cdot u \notag \\
&+\frac{ig_{A}}{8m^{3}}\upsilon \cdot u (\upsilon \cdot D)^{2}  S \cdot D +\frac{g_{A}^{2}}{8m^{3}}\upsilon \cdot u  \{\upsilon \cdot D, S \cdot u \}  S \cdot D\notag \\
&-\frac{ig_{A}^{3}}{8m^{3}}\upsilon \cdot u (S \cdot u )^{2} S \cdot D +\frac{ig_{A}}{4m^{2}}\upsilon \cdot u C_{(2)} S \cdot D  , \\
-\frac{1}{m^{2}}S \cdot D C_{(2)} S \cdot D&=\frac{1}{m^{2}}S \cdot D \{b_{1}\left \langle \chi_{+} \right \rangle +b_{2} \widetilde{\chi}_{+}+b_{3}u^{2}+b_{4}\left \langle u^{2} \right \rangle \notag \\
&+2b_{7}[S^{\mu}, S^{\nu}][u_{\mu},u_{\nu}]-2ib_{8}[S_{\mu}, S_{\nu}]f_{+}^{\mu\nu}\}S \cdot D \notag \\
&-4S \cdot D\{2b_{5}(\upsilon \cdot u)^{2}+b_{6}\left \langle (\upsilon \cdot u)^{2} \right \rangle\} S \cdot D ,\\
\frac{g_{A}^{2}}{16m^{2}}\upsilon \cdot u C_{(2)} \upsilon \cdot u&=-\frac{g_{A}^{2}}{16m^{2}}\upsilon \cdot u \{b_{1}\left \langle \chi_{+} \right \rangle +b_{2} \widetilde{\chi}_{+}+b_{3}u^{2}+b_{4}\left \langle u^{2} \right \rangle \notag \\
&+2b_{7}[S^{\mu}, S^{\nu}][u_{\mu},u_{\nu}]-2ib_{8}[S_{\mu}, S_{\nu}]f_{+}^{\mu\nu}\}\upsilon \cdot u \notag \\
&+\frac{g_{A}^{2}}{4}\upsilon \cdot u\{2b_{5}(\upsilon \cdot u)^{2}+b_{6}\left \langle (\upsilon \cdot u)^{2} \right \rangle\} \upsilon \cdot u,\\
\frac{ig_{A}}{2m^{2}}S \cdot D C_{(2)} \upsilon \cdot u &=-\frac{ig_{A}}{2m^{2}}S \cdot D \{b_{1}\left \langle \chi_{+} \right \rangle +b_{2} \widetilde{\chi}_{+}+b_{3}u^{2}+b_{4}\left \langle u^{2} \right \rangle \notag \\
&+2b_{7}[S^{\mu}, S^{\nu}][u_{\mu},u_{\nu}]-2ib_{8}[S_{\mu}, S_{\nu}]f_{+}^{\mu\nu}\}\upsilon \cdot u \notag \\
&+2ig_{A}S \cdot D\{2b_{5}(\upsilon \cdot u)^{2}+b_{6}\left \langle (\upsilon \cdot u)^{2} \right \rangle\} \upsilon \cdot u ,\\
\frac{ig_{A}}{2m^{2}}\upsilon \cdot u C_{(2)} S \cdot D &=-\frac{ig_{A}}{2m^{2}}\upsilon \cdot u \{b_{1}\left \langle \chi_{+} \right \rangle +b_{2} \widetilde{\chi}_{+}+b_{3}u^{2}+b_{4}\left \langle u^{2} \right \rangle \notag \\
&+2b_{7}[S^{\mu}, S^{\nu}][u_{\mu},u_{\nu}]-2ib_{8}[S_{\mu}, S_{\nu}]f_{+}^{\mu\nu}\} S \cdot D \notag \\
&+2ig_{A}\upsilon \cdot u \{2b_{5}(\upsilon \cdot u)^{2}+b_{6}\left \langle (\upsilon \cdot u)^{2} \right \rangle\} S \cdot D,\\
\gamma_{0}\mathcal{B}^{(2)\dagger}\gamma_{0}\mathcal{C}^{(0)-1}\mathcal{B}^{(2)}&=\frac{2}{m}(\upsilon^{\mu}S^{\nu}-\upsilon^{\nu}S^{\mu})^{2}\{b_{7}^{2}([u_{\mu},u_{\nu}])^{2}-b_{8}^{2}(f_{+\mu\nu})^{2} \}\notag \\
&-\frac{2i}{m}b_{7}b_{8}(\upsilon^{\mu}S^{\nu}-\upsilon^{\nu}S^{\mu})^{2}\{[u_{\mu},u_{\nu}],f_{+}^{\mu\nu}\},\\
\gamma_{0}\mathcal{B}^{(2)\dagger}\gamma_{0}\mathcal{C}^{(1)-1}\mathcal{B}^{(1)}&=\frac{1}{m^{2}}b_{7}(\upsilon^{\mu}S^{\nu}-\upsilon^{\nu}S^{\mu})[u_{\mu},u_{\nu}] (\upsilon \cdot D -ig_{A}S \cdot u)S \cdot D \notag \\
&-\frac{ig_{A}}{4m^{2}}b_{7}(\upsilon^{\mu}S^{\nu}-\upsilon^{\nu}S^{\mu})[u_{\mu},u_{\nu}] (\upsilon \cdot D-ig_{A} S \cdot u) \upsilon \cdot u \notag \\
&- \frac{i}{m^{2}}b_{8}(\upsilon^{\mu}S^{\nu}-\upsilon^{\nu}S^{\mu})f_{+\mu\nu} (\upsilon \cdot D-ig_{A}S \cdot u) S \cdot D \notag \\
&-\frac{g_{A}}{4m^{2}}b_{8}(\upsilon^{\mu}S^{\nu}-\upsilon^{\nu}S^{\mu})f_{+\mu\nu} (\upsilon \cdot D-ig_{A}S \cdot u) \upsilon \cdot u , \\
\gamma_{0}\mathcal{B}^{(3)\dagger}\gamma_{0}\mathcal{C}^{(0)-1}\mathcal{B}^{(1)}&=4\epsilon_{\mu\nu\lambda\tau}(c_{9}\{u^{\mu},\{u^{\nu},\upsilon \cdot u\}\}+c_{10}u^{\mu}\left \langle u^{\nu}\upsilon \cdot u\right \rangle )(\upsilon^{\lambda}S^{\tau}-\upsilon^{\tau}S^{\lambda})S \cdot D \notag \\
&+4(c_{13}\{u^{\mu},h^{\nu\rho}\}+c_{14}\left \langle u^{\mu}h^{\nu\rho} \right \rangle )(\upsilon_{\mu}S_{\nu}-\upsilon_{\nu}S_{\mu})\upsilon_{\rho}S \cdot D \notag \\
&-4ic_{24}\epsilon_{\nu\rho\lambda\tau}[\upsilon \cdot u,f^{\nu \rho}_{+}](\upsilon^{\lambda}S^{\tau}-\upsilon^{\tau}S^{\lambda}) S \cdot D \notag \\
&+4(c_{29}\{u^{\mu},f^{\nu\rho}_{-}\}+c_{30} \left \langle u^{\mu}f^{\nu\rho}_{-}\right  \rangle)(\upsilon_{\mu}S_{\nu}-\upsilon_{\mu}S_{\nu})\upsilon_{\rho}S \cdot D \notag \\
&+\frac{i}{m}(c_{1}\{u_{\mu},\{u^{\mu},\upsilon \cdot u\}\} -2ic_{11}[u_{\mu},h^{\mu \nu}]S_{\nu}+c_{15}\{\upsilon \cdot u ,\widetilde{\chi}_{+}\}  \notag \\
&+c_{16}\upsilon \cdot u \left \langle \chi_{+} \right \rangle+c_{17}\left \langle \upsilon \cdot  u \widetilde{\chi}_{+} \right \rangle +ic_{18}[\upsilon \cdot D, \widetilde{\chi}_{-}] +c_{2}[u_{\mu},[u^{\mu}, \upsilon \cdot u]] \notag \\
&+ic_{19}\left \langle [\upsilon \cdot D, \chi_{-}] \right \rangle+2c_{20}[\widetilde{\chi}_{-}, S \cdot u]+ic_{21}[u_{\mu},f_{+}^{\mu\nu}]\upsilon_{\nu})S \cdot D \notag \\
&-\frac{2i}{m}\epsilon_{\mu\nu\rho\tau} (c_{22}\{u^{\mu},f_{+}^{\nu\rho}\}+c_{23}\left \langle u^{\mu}f_{+}^{\nu\rho} \right \rangle)S^{\tau} S \cdot D \notag \\
&+\frac{2}{m}c_{26}[u_{\mu},f^{\mu \nu}_{-}]S_{\nu} S \cdot D -\frac{i}{m}c_{31}\upsilon_{\nu}[D_{\mu},f^{\mu\nu}_{-}]S \cdot D\notag \\
&+\frac{i}{m}\epsilon_{\mu\nu\rho\tau} (c_{27}\{u^{\mu},f_{-}^{\nu\rho}\}+c_{28}\left \langle u^{\mu}f_{-}^{\nu\rho} \right  \rangle)\upsilon^{\tau} S \cdot D \notag \\
&+\frac{i}{m}(c_{3}\upsilon \cdot u \left \langle u^{2} \right \rangle+c_{4}u_{\mu} \left \langle u^{\mu}  \upsilon \cdot u \right \rangle ) S \cdot D \notag \\
&+\frac{2}{m}\epsilon_{\mu\nu\rho\tau}(c_{7}\{[u^{\mu},u^{\nu}],u^{\rho}\}+c_{8}\left \langle [u^{\mu},u^{\nu}]u^{\rho}\right \rangle )S^{\tau} S \cdot D \notag \\
&-8mc_{12}[S \cdot u ,h^{\nu\rho}]\upsilon_{\nu} \upsilon_{\rho}S \cdot D +4imc_{32} [\upsilon \cdot D,f^{\mu\nu}_{-}]\upsilon_{\mu}\upsilon_{\nu} S \cdot D  \notag \\
&-4im(4c_{5}(\upsilon \cdot u)^{3}+c_{6}\upsilon \cdot u \left \langle (\upsilon \cdot u)^{2} \right \rangle)S \cdot D \notag \\
&-ig_{A}\epsilon_{\mu\nu\lambda\tau}(c_{9}\{u^{\mu},\{u^{\nu},\upsilon \cdot u\}\}+c_{10}u^{\mu}\left \langle u^{\nu}\upsilon \cdot u\right \rangle)(\upsilon^{\lambda}S^{\tau}-\upsilon^{\tau}S^{\lambda})\upsilon \cdot u \notag \\
&-ig_{A}(c_{13}\{u^{\mu},h^{\nu\rho}\}+c_{14}\left \langle u^{\mu}h^{\nu\rho} \right \rangle )(\upsilon_{\mu}S_{\nu}-\upsilon_{\nu}S_{\mu})\upsilon_{\rho}\upsilon \cdot u \notag \\
&-g_{A}c_{24}\epsilon_{\nu\rho\lambda\tau}[\upsilon \cdot u,f^{\nu \rho}_{+}](\upsilon^{\lambda}S^{\tau}-\upsilon^{\tau}S^{\lambda})\upsilon \cdot u \notag \\
&-ig_{A}(c_{29}\{u^{\mu},f^{\nu\rho}_{-}\}+c_{30}\left \langle u^{\mu}f^{\nu\rho}_{-}\right  \rangle)(\upsilon_{\mu}S_{\nu}-\upsilon_{\mu}S_{\nu})\upsilon_{\rho}\upsilon \cdot u \notag \\
&+\frac{g_{A}}{4m}(c_{1}\{u_{\mu},\{u^{\mu},\upsilon \cdot u\}\}-2ic_{11}[u_{\mu},h^{\mu \nu}]S_{\nu}+c_{15}\{\upsilon \cdot u ,\widetilde{\chi}_{+}\} \notag \\
&+c_{16}\upsilon \cdot u \left \langle \chi_{+} \right \rangle+c_{17}\left \langle \upsilon \cdot  u \widetilde{\chi}_{+} \right \rangle+ic_{18}[\upsilon \cdot D, \widetilde{\chi}_{-}]+c_{2}[u_{\mu},[u^{\mu}, \upsilon \cdot u]]\notag \\
&+ic_{19}\left \langle [\upsilon \cdot D, \chi_{-}] \right \rangle+2c_{20}[\widetilde{\chi}_{-}, S \cdot u]+ic_{21}[u_{\mu},f_{+}^{\mu\nu}]\upsilon_{\nu})\upsilon \cdot u \notag \\
&-\frac{g_{A}}{2m}\epsilon_{\mu\nu\rho\tau} (c_{22}\{u^{\mu},f_{+}^{\nu\rho}\}+c_{23}\left \langle u^{\mu}f_{+}^{\nu\rho} \right \rangle)S^{\tau} \upsilon \cdot u \notag \\
&-\frac{ig_{A}}{2m}c_{26}[u_{\mu},f^{\mu \nu}_{-}]S_{\nu} \upsilon \cdot u +\frac{g_{A}}{4m}c_{3}\upsilon \cdot u \left \langle u^{2} \right \rangle \upsilon \cdot u\notag \\
&+\frac{g_{A}}{4m}\epsilon_{\mu\nu\rho\tau}(c_{27} \{u^{\mu},f_{-}^{\nu\rho}\}+c_{28}\left \langle u^{\mu}f_{-}^{\nu\rho} \right  \rangle )\upsilon^{\tau} \upsilon \cdot u \notag \\
&-\frac{g_{A}}{4m}c_{31}\upsilon_{\nu}[D_{\mu},f^{\mu\nu}_{-}] \upsilon \cdot u +\frac{g_{A}}{4m}c_{4}u_{\mu} \left \langle u^{\mu}  \upsilon \cdot u \right \rangle \upsilon \cdot u\notag \\
&-\frac{ig_{A}}{2m}\epsilon_{\mu\nu\rho\tau}(c_{7}\{[u^{\mu},u^{\nu}],u^{\rho}\} +c_{8} \left \langle [u^{\mu},u^{\nu}]u^{\rho}\right \rangle)S^{\tau} \upsilon \cdot u \notag \\
&+2img_{A}c_{12}[S \cdot u ,h^{\nu\rho}]\upsilon_{\nu} \upsilon_{\rho} \upsilon \cdot u +mg_{A}c_{32}[\upsilon \cdot D,f^{\mu\nu}_{-}]\upsilon_{\mu}\upsilon_{\nu} \upsilon \cdot u \notag \\
&-4mg_{A}c_{5}(\upsilon \cdot u)^{4} -mg_{A}c_{6} \upsilon \cdot u \left \langle (\upsilon \cdot u)^{2} \right \rangle \upsilon \cdot u.  
\end{align}

\section{\label{app:C}The $\mathcal{O}(q^4)$ operators}
\begin{longtable}{ccc}
\caption{Terms in the $O(q^{4})$ relativistic and non-relativistic Lagrangians.}
\label{tab:p4}\\
\toprule
\midrule
$i$ & $O_{i}^{(4)}$ & $\hat{O}_{i}^{(4)}$\\
\bottomrule
\endhead
\midrule
\bottomrule
\endfoot
\bottomrule
\bottomrule
\endlastfoot
$1$  & $\{u^{\mu},\{\{u_{\mu},u^{\nu}\},u_{\nu}\}\}$  & $\{u^{\mu},\{\{u_{\mu},u^{\nu}\},u_{\nu}\}\}$ \\
$2$  & $\{u^{\mu},[[u_{\mu},u^{\nu}],u_{\nu}]\}$ & $\{u^{\mu},[[u_{\mu},u^{\nu}],u_{\nu}]\}$  \\
$3$  & $[u^{\mu},\{[u_{\mu},u^{\nu}],u_{\nu}\}]$   & $[u^{\mu},\{[u_{\mu},u^{\nu}],u_{\nu}\}] $ \\
$4$  & $ \left \langle \{u^{\mu},\{u_{\mu},u^{\nu}\} \} u_{\nu} \right \rangle $  & $ \left \langle \{u^{\mu},\{u_{\mu},u^{\nu}\} \} u_{\nu} \right \rangle $ \\
$5$  & $\left \langle  [u^{\mu},[u_{\mu},u^{\nu}]] u_{\nu} \right \rangle  $   & $ \left \langle  [u^{\mu},[u_{\mu},u^{\nu}]] u_{\nu} \right \rangle  $  \\
$6$  & $u^{\mu}\left \langle u^{2}u_{\mu} \right \rangle $  & $u^{\mu} \left \langle u^{2}u_{\mu} \right \rangle $\\
$7$  & $u^{\mu}u_{\mu}\left \langle u^{2} \right \rangle$   & $u^{2} \left \langle u^{2} \right \rangle$  \\
$8$  & $u^{\mu}u^{\nu}\left \langle u_{\mu}u_{\nu} \right \rangle $   & $u^{\mu}u^{\nu} \left \langle u_{\mu}u_{\nu} \right \rangle $  \\
$9$  & $i\{u^{\mu},\{u_{\mu},[u^{\nu},u^{\lambda}]\}\}\sigma_{\nu\lambda}$   & $2\{u^{\mu},\{u_{\mu},[u^{\nu},u^{\lambda}]\}\} [S_{\nu}, S_{\lambda}]$  \\
$10$  & $i[u^{\mu},[u_{\mu},[u^{\nu},u^{\lambda}]]]\sigma_{\nu\lambda} $  & $2[u^{\mu},[u_{\mu},[u^{\nu},u^{\lambda}]]][S_{\nu}, S_{\lambda}]$ \\
$11$  & $i[u^{\nu},\{u_{\mu},\{u^{\mu},u^{\lambda}\}\}]\sigma_{\nu\lambda} $   & $2[u^{\nu},\{u_{\mu},\{u^{\mu},u^{\lambda}\}\}][S_{\nu}, S_{\lambda}]$ \\
$12$  & $i[u^{\nu},[u_{\mu},[u^{\mu},u^{\lambda}]]]\sigma_{\nu\lambda} $  & $2[u^{\nu},[u_{\mu},[u^{\mu},u^{\lambda}]]][S_{\nu}, S_{\lambda}]$ \\
$13$  & $i\left \langle [u_{\mu},u_{\nu}]u^{2} \right \rangle \sigma^{\mu\nu}$  & $2[S^{\mu}, S^{\nu}]\left \langle [u_{\mu},u_{\nu}]u^{2} \right \rangle$\\
$14$  & $i\left \langle [u_{\mu},\{u^{\lambda},u_{\nu}\}]u_{\lambda} \right \rangle \sigma^{\mu\nu} $   & $2[S^{\mu}, S^{\nu}]\left \langle [u_{\mu},\{u^{\lambda},u_{\nu}\}]u_{\lambda} \right \rangle  $ \\
$15$  & $iu^{\mu}\left \langle u_{\mu}[u_{\nu},u_{\lambda}] \right \rangle \sigma^{\nu\lambda}$  & $2[S^{\nu}, S^{\lambda}]u^{\mu}\left \langle u_{\mu}[u_{\nu},u_{\lambda}] \right \rangle$ \\
$16$  & $i[u^{\mu},u^{\nu}]\left \langle u^{2} \right \rangle  \sigma_{\mu\nu}$   & $2[u^{\mu},u^{\nu}][S_{\mu}, S_{\nu}]\left \langle u^{2} \right \rangle  $ \\
$17$  & $i[u^{\mu},u^{\lambda}]\left \langle u^{\nu}u_{\lambda} \right \rangle \sigma_{\mu\nu} $   & $2[u^{\mu},u^{\lambda}][S_{\mu}, S_{\nu}]\left \langle u^{\nu}u_{\lambda} \right \rangle $ \\
$18$  & $\{u^{\mu},\{u_{\mu},\{u^{\nu},u^{\lambda}\}\}\}D_{\nu\lambda} + h.c. $   & $-8m^{2}\{u^{\mu},\{u_{\mu},(\upsilon \cdot u)^{2} \}\} $ \\
$19$  & $[u^{\mu},[u_{\mu},\{u^{\nu},u^{\lambda}\}]]D_{\nu\lambda} + h.c.$  & $-8m^{2}[u^{\mu},[u_{\mu},(\upsilon \cdot u)^{2}]]$\\
$20$  & $\{u^{\nu},\{u_{\mu},\{u^{\mu},u^{\lambda}\}\}\}D_{\nu\lambda}+h.c. $   & $-4m^{2}\{\upsilon \cdot u,\{u_{\mu},\{u^{\mu},\upsilon \cdot u\}\}\}$ \\
$21$  & $\{u^{\nu},[u_{\mu},[u^{\mu},u^{\lambda}]]\}D_{\nu\lambda}+h.c.$  & $-4m^{2}\{\upsilon \cdot u,[u_{\mu},[u^{\mu},\upsilon \cdot u]]\}$ \\
$22$  & $ \left \langle \{u_{\mu},\{u^{\lambda},u_{\nu}\}\}u_{\lambda} \right \rangle D^{\mu\nu} + h.c.  $   & $-4m^{2} \left \langle \{\upsilon \cdot u,\{u^{\lambda},\upsilon \cdot u\}\}u_{\lambda} \right \rangle   $ \\
$23$  & $ \left \langle \{u_{\mu},u_{\nu}\}u^{2} \right \rangle D^{\mu\nu}+ h.c.$   & $-8m^{2} \left \langle (\upsilon \cdot u)^{2}u^{2} \right \rangle $  \\
$24$  & $u^{\mu}\left \langle u_{\nu}u^{2} \right \rangle {D_{\mu}}^{\nu}+ h.c. $  & $-4m^{2}\upsilon \cdot u \left \langle \upsilon \cdot u u^{2} \right \rangle $ \\
$25$  & $u^{\lambda} \left \langle \{u_{\mu},u_{\nu}\}u_{\lambda} \right \rangle D^{\mu\nu}+ h.c.$  & $ -8m^{2}u^{\lambda} \left \langle (\upsilon \cdot u)^{2}u_{\lambda} \right \rangle $ \\
$26$  & $u^{2}\left \langle u_{\nu}u_{\lambda} \right \rangle D^{\nu\lambda}+ h.c.  $   & $-4m^{2}u^{2}\left \langle (\upsilon \cdot u)^{2} \right \rangle  $ \\
$27$  & $ \{u^{\mu},u^{\nu}\} \left \langle u^{2} \right \rangle D_{\mu\nu}+ h.c.  $  & $ -8m^{2}(\upsilon \cdot u)^{2} \left \langle u^{2} \right \rangle  $ \\
$28$  & $\{u^{\mu},u^{\nu}\} \left \langle u_{\nu} u_{\lambda}\right \rangle {D_{\mu}}^{\lambda}+ h.c.  $   & $-4m^{2}\{\upsilon \cdot u ,u^{\nu}\} \left \langle u_{\nu} \upsilon \cdot u\right \rangle  $ \\
$29$  & $ i\{u^{\lambda},\{u^{\rho},[u^{\mu},u^{\nu}]\}\}\sigma_{\mu\nu}D_{\lambda\rho}+h.c.$  & $ -8m^{2}\{\upsilon \cdot u ,\{\upsilon \cdot u ,[u^{\mu},u^{\nu}]\}\}[S_{\mu},S_{\nu}]$ \\
$30$  & $i[u^{\lambda},[u^{\rho},[u^{\mu},u^{\nu}]]]\sigma_{\mu\nu}D_{\lambda\rho} +h.c. $  & $-8m^{2}[\upsilon \cdot u,[\upsilon \cdot u ,[u^{\mu},u^{\nu}]]] [S_{\mu},S_{\nu}] $ \\
$31$  & $i[u^{\mu},\{u^{\lambda},\{u^{\rho},u^{\nu}\}\}]\sigma_{\mu\nu}D_{\lambda\rho} +h.c. $   & $-8m^{2}[u^{\mu},\{\upsilon \cdot u,\{\upsilon \cdot u,u^{\nu}\}\}] [S_{\mu},S_{\nu}]$ \\
$32$  & $ i[u^{\mu},[u^{\lambda},[u^{\rho},u^{\nu}]]]\sigma_{\mu\nu}D_{\lambda\rho} +h.c.  $  &  $ -8m^{2}[u^{\mu},[\upsilon \cdot u,[\upsilon \cdot u,u^{\nu}]]] [S_{\mu},S_{\nu}]$ \\
$33$  & $  i \left \langle \{[u_{\mu},u_{\nu}],u_{\lambda}\}u_{\rho} \right \rangle \sigma^{\mu\nu} D^{\lambda\rho}+h.c.$   & $ -8m^{2} \left \langle \{[u_{\mu},u_{\nu}],\upsilon \cdot u\}\upsilon \cdot u \right \rangle [S^{\mu},S^{\nu}] $ \\
$34$  & $ i\left \langle [u_{\mu},\{u_{\lambda},u_{\nu}\}]u_{\rho} \right \rangle \sigma^{\mu\nu} D^{\lambda\rho}+h.c.$   & $-8m^{2}  \left \langle [u_{\mu},\{\upsilon \cdot u,u_{\nu}\}]\upsilon \cdot u \right \rangle [S^{\mu},S^{\nu}] $ \\
$35$  & $iu^{\mu} \left \langle [u_{\nu},u_{\lambda}]u_{\rho} \right \rangle \sigma^{\nu\lambda} {D_{\mu}}^{\rho} +h.c. $   & $-8m^{2} \upsilon \cdot u  \left \langle [u_{\nu},u_{\lambda}]\upsilon \cdot u  \right \rangle [S^{\nu},S^{\lambda}] $ \\
$36$  & $i[u^{\mu},u^{\nu}]\left \langle u_{\lambda}u_{\rho} \right \rangle \sigma_{\mu\nu} D^{\lambda\rho} +h.c. $  & $-8m^{2}[u^{\mu},u^{\nu}] [S_{\mu},S_{\nu}]\left \langle (\upsilon \cdot u)^{2} \right \rangle $ \\
$37$  & $i[u^{\mu},u^{\nu}]\left \langle u_{\lambda}u_{\rho} \right \rangle{\sigma_{\mu}}^{\lambda} {D_{\nu}}^{\rho}+ h.c.  $   &  $-8m^{2}[u^{\mu},\upsilon \cdot u ] [S_{\mu},S^{\lambda}]\left \langle u_{\lambda}\upsilon \cdot u\right \rangle  $  \\
$38$  & $ \{u^{\mu},\{u^{\nu},\{u^{\lambda},u^{\rho}\}\}\}D_{\mu\nu\lambda\rho}+ h.c. $   & $ 384m^{4}(\upsilon \cdot u)^{4} $ \\
$39$  & $ \left \langle \{u_{\mu},\{u_{\nu},u_{\lambda}\}\}u_{\rho} \right \rangle D^{\mu\nu\lambda\rho}+ h.c.  $  & $192m^{4} \left \langle (\upsilon \cdot u)^{4} \right \rangle  $  \\
$40$  & $u^{\mu} \left \langle \{u_{\nu},u_{\lambda}\}u_{\rho} \right \rangle {D_{\mu}}^{ \nu\lambda\rho} + h.c.  $   & $96m^{4}\upsilon \cdot u \left \langle (\upsilon \cdot u)^{3}  \right \rangle   $ \\
$41$  & $ \{u^{\mu},u^{\nu}\} \left \langle u_{\lambda}u_{\rho} \right \rangle {D_{\mu\nu}}^{\lambda\rho}+ h.c. $   & $96m^{4}(\upsilon \cdot u)^{2} \left \langle (\upsilon \cdot u)^{2}  \right \rangle  $ \\
$42$  & $ \epsilon^{\mu\nu\lambda\rho}[[u_{\mu},u_{\nu}],f_{-\lambda\rho} ]  $   & $ \epsilon^{\mu\nu\lambda\rho}[[u_{\mu},u_{\nu}],f_{-\lambda\rho} ] $ \\
$43$  & $\epsilon^{\mu\nu\lambda\rho}u_{\mu} \left \langle u_{\nu}f_{-\lambda\rho} \right \rangle  $   & $\epsilon^{\mu\nu\lambda\rho} u_{\mu} \left \langle u_{\nu}f_{-\lambda\rho} \right \rangle  $ \\
$44$  & $ \{[u^{\mu} ,u^{\nu}],{f_{-\mu}}^{\lambda}\}\gamma_{5}\gamma_{\nu}D_{\lambda}  +h.c.$   & $ 4im\{[u^{\mu} ,S \cdot u ],{f_{-\mu}}^{\lambda}\}\upsilon_{\lambda}$ \\
$45$  & $  [\{u^{\mu} ,u^{\nu}\},{f_{-\mu}}^{\lambda}]\gamma_{5}\gamma_{\nu}D_{\lambda}  +h.c. $   & $4im  [\{u^{\mu} ,S \cdot u \},{f_{-\mu}}^{\lambda}]\upsilon_{\lambda}  $ \\
$46$  & $\{[u^{\mu} ,u^{\nu}],{f_{-\mu}}^{\lambda}\}\gamma_{5}\gamma_{\lambda}D_{\nu} +h.c. $   & $4im \{[u^{\mu} ,\upsilon \cdot u ],{f_{-\mu}}^{\lambda}\}S_{\lambda} $ \\
$47$  & $   [\{u^{\mu} ,u^{\nu}\},{f_{-\mu}}^{\lambda}]\gamma_{5}\gamma_{\lambda}D_{\nu} +h.c. $   & $4im [\{u^{\mu} ,\upsilon \cdot u \},{f_{-\mu}}^{\lambda}]S_{\lambda} $ \\
$48$  & $ \{u^{\mu},[ u^{\lambda},{f_{-\mu}}^{\nu}]\}\gamma_{5}\gamma_{\nu}D_{\lambda}  +h.c.$   & $4im \{u^{\mu},[ \upsilon \cdot u ,{f_{-\mu}}^{\nu}]\}S_{\nu}$ \\
$49$  & $ \{u^{\mu},[ u^{\lambda},{f_{-\mu}}^{\nu}]\}\gamma_{5}\gamma_{\lambda}D_{\nu}  +h.c. $  & $4im \{u^{\mu},[ S \cdot u ,{f_{-\mu}}^{\nu}]\} \upsilon_{\nu}$ \\
$50$  & $\{u^{\mu},[u_{\mu},f_{-}^{\nu\lambda}]\}\gamma_{5}\gamma_{\nu}D_{\lambda} +h.c.  $   & $4im\{u^{\mu},[u_{\mu},f_{-}^{\nu\lambda}]\}S_{\nu}\upsilon_{\lambda}  $ \\
$51$  & $ \left \langle [u_{\mu},u^{\lambda}]f_{-\nu\lambda} \right \rangle \gamma_{5}\gamma^{\mu} D^{\nu}  +h.c.  $   & $4im \left \langle [S \cdot u ,u^{\lambda}]f_{-\nu\lambda} \right \rangle \upsilon^{\nu}  $ \\
$52$  & $\left \langle [u_{\nu},u^{\lambda}]f_{-\mu\lambda} \right \rangle \gamma_{5}\gamma^{\mu} D^{\nu}+h.c.  $   & $4im \left \langle [\upsilon \cdot u,u^{\lambda}]f_{-\mu\lambda} \right \rangle  S^{\mu}$  \\
$53$  & $  \{u^{\mu},[ u_{\mu},h^{\nu\lambda}]\}\gamma_{5}\gamma_{\nu}D_{\lambda}  +h.c.$   & $ 4im \{u^{\mu},[ u_{\mu},h^{\nu\lambda}]\}S_{\nu}\upsilon_{\lambda}$ \\
$54$  & $\{[u^{\mu},u^{\nu}],{h_{\mu}}^{\lambda}\}\gamma_{5}\gamma_{\nu}D_{\lambda}  +h.c. $   & $4im \{[u^{\mu},S \cdot u ],{h_{\mu}}^{\lambda}\}\upsilon_{\lambda}$  \\
$55$  & $ [\{u^{\mu},u^{\nu}\},{h_{\mu}}^{\lambda}]\gamma_{5}\gamma_{\nu}D_{\lambda} +h.c.  $   & $4im [\{u^{\mu},S \cdot u \},{h_{\mu}}^{\lambda}]\upsilon_{\lambda}  $\\
$56$  & $ \{[u^{\mu}, u^{\nu}],{h_{\mu}}^{\lambda}\}\gamma_{5}\gamma_{\lambda}D_{\nu}  +h.c. $   & $4im \{[u^{\mu}, \upsilon \cdot u ],{h_{\mu}}^{\lambda}\}S_{\lambda} $\\
$57$  & $ [\{u^{\mu} ,u^{\nu}\},{h_{\mu}}^{\lambda}]\gamma_{5}\gamma_{\lambda}D_{\nu} +h.c.  $  &  $4im [\{u^{\mu} ,\upsilon \cdot u\},{h_{\mu}}^{\lambda}] S_{\lambda} $\\
$58$  & $ \{u^{\mu},[u^{\lambda}, {h_{\mu}}^{\nu}]\}\gamma_{5}\gamma_{\nu}D_{\lambda}  +h.c.  $   & $4im \{u^{\mu},[\upsilon \cdot u, {h_{\mu}}^{\nu}]\}S_{\nu} $ \\
$59$  & $ \{u^{\mu},[u^{\lambda}, {h_{\mu}}^{\nu}]\}\gamma_{5}\gamma_{\lambda}D_{\nu}  +h.c.  $ & $4im  \{u^{\mu},[S \cdot u, {h_{\mu}}^{\nu}]\}\upsilon_{\nu}  $  \\
$60$  & $\left \langle [u_{\mu},u^{\lambda}]h_{\nu\lambda} \right \rangle \gamma_{5}\gamma^{\mu}  D^{\nu}+h.c.  $   & $4im \left \langle [S \cdot u ,u^{\lambda}]h_{\nu\lambda} \right \rangle \upsilon^{\nu} $  \\
$61$  & $ \left \langle [u_{\nu},u^{\lambda}]h_{\mu\lambda} \right \rangle \gamma_{5}\gamma^{\mu}  D^{\nu}+h.c.  $   & $4im  \left \langle [\upsilon \cdot u ,u^{\lambda}]h_{\mu\lambda} \right \rangle S^{\mu} $ \\
$62$  & $\epsilon^{\mu\nu\lambda\rho}[[u_{\mu},u_{\nu}],{f_{-\lambda}}^{\sigma}]D_{\rho\sigma}+h.c. $   & $-4m^{2}\epsilon^{\mu\nu\lambda\rho}[[u_{\mu},u_{\nu}],{f_{-\lambda}}^{\sigma}]\upsilon_{\rho}\upsilon_{\sigma}$ \\
$63$  & $ \epsilon^{\mu\nu\lambda\rho}\{\{u_{\mu},u^{\sigma}\},f_{-\nu\lambda}\}D_{\rho\sigma} +h.c.$   & $ -4m^{2}\epsilon^{\mu\nu\lambda\rho}\{\{u_{\mu},\upsilon \cdot u \},f_{-\nu\lambda}\}\upsilon_{\rho}$ \\ 
$64$  & $\epsilon^{\mu\nu\lambda\rho}[[u_{\mu},u^{\sigma}],f_{-\nu\lambda}]D_{\rho\sigma} +h.c.  $   & $-4m^{2}\epsilon^{\mu\nu\lambda\rho}[[u_{\mu},\upsilon \cdot u],f_{-\nu\lambda}]\upsilon_{\rho} $ \\
$65$  & $\epsilon^{\mu\nu\lambda\rho}\{u_{\mu},\{f_{-\nu\lambda},u^{\sigma}\}\}D_{\rho\sigma} +h.c. $  & $-4m^{2}\epsilon^{\mu\nu\lambda\rho}\{u_{\mu},\{f_{-\nu\lambda},\upsilon \cdot u \}\}\upsilon_{\rho} $ \\
$66$  & $\epsilon^{\mu\nu\lambda\rho} \left \langle \{u_{\nu},u_{\sigma}\}f_{-\lambda\rho} \right \rangle {D_{\mu}}^{\sigma} +h.c.  $  & $-4m^{2}\epsilon^{\mu\nu\lambda\rho}\upsilon_{\mu} \left \langle \{u_{\nu},\upsilon \cdot u \}f_{-\lambda\rho} \right \rangle  $ \\
$67$  & $ \epsilon^{\mu\nu\lambda\rho}u_{\mu} \left \langle u_{\nu}{f_{-\lambda}}^{\sigma} \right \rangle D_{\rho\sigma} +h.c.$   & $ -4m^{2}\epsilon^{\mu\nu\lambda\rho}u_{\mu}\upsilon_{\rho}\upsilon_{\sigma}  \left \langle u_{\nu}{f_{-\lambda}}^{\sigma} \right \rangle $ \\
$68$  & $\epsilon^{\mu\nu\lambda\rho}u_{\mu} \left \langle u^{\sigma}f_{-\nu\lambda} \right \rangle D_{\rho\sigma} +h.c. $  & $-4m^{2}\epsilon^{\mu\nu\lambda\rho}u_{\mu}\upsilon_{\rho} \left \langle \upsilon \cdot u f_{-\nu\lambda} \right \rangle $ \\
$69$  & $ \epsilon^{\mu\nu\lambda\rho}u^{\sigma}  \left \langle u_{\mu}f_{-\nu\lambda} \right \rangle D_{\rho\sigma} +h.c. $   & $ -4m^{2}\epsilon^{\mu\nu\lambda\rho}\upsilon \cdot u \upsilon_{\rho}  \left \langle u_{\mu}f_{-\nu\lambda} \right \rangle$ \\
$70$  & $\epsilon^{\mu\nu\lambda\rho}f_{-\mu\nu} \left \langle u_{\rho}u_{\sigma} \right \rangle {D_{\lambda}}^{\sigma} +h.c.  $  & $-4m^{2}\epsilon^{\mu\nu\lambda\rho}f_{-\mu\nu}\upsilon_{\lambda} \left \langle u_{\rho} \upsilon \cdot u \right \rangle  $ \\
$71$  & $ \epsilon^{\mu\nu\lambda\rho}[[u_{\mu},u_{\nu}],{h_{\lambda}}^{\sigma}]D_{\rho\sigma} +h.c. $   &  $-4m^{2} \epsilon^{\mu\nu\lambda\rho}[[u_{\mu},u_{\nu}],{h_{\lambda}}^{\sigma}]\upsilon_{\rho}\upsilon_{\sigma}$  \\
$72$  & $\epsilon^{\mu\nu\lambda\rho}u_{\mu} \left \langle u_{\lambda}h_{\rho\sigma} \right \rangle {D_{\nu}}^{\sigma}+h.c.  $   & $-4m^{2}\epsilon^{\mu\nu\lambda\rho}u_{\mu}\upsilon_{\nu}\upsilon^{\sigma} \left \langle u_{\lambda}h_{\rho\sigma} \right \rangle  $  \\
$73$  & $[\{u^{\mu},u^{\nu}\},f_{-}^{\lambda\rho}]\gamma_{5}\gamma_{\lambda}D_{\mu\nu\rho} +h.c. $   & $-48im^{3}[(\upsilon \cdot u)^{2},f_{-}^{\lambda\rho}] S_{\lambda}\upsilon_{\rho} $ \\
$74$  & $[\{u^{\mu},u^{\nu}\},h^{\lambda\rho}]\gamma_{5}\gamma_{\lambda}D_{\mu\nu\rho} +h.c. $   & $-48im^{3}[(\upsilon \cdot u)^{2},h^{\lambda\rho}] S_{\lambda}\upsilon_{\rho} $ \\
$75$  & $  \{[u^{\mu},u^{\nu}],h^{\lambda\rho}\}\gamma_{5}\gamma_{\mu}D_{\nu\lambda\rho} +h.c.$  & $-24im^{3} \{[S \cdot u ,\upsilon \cdot u ],h^{\lambda\rho}\}\upsilon_{\lambda}\upsilon_{\rho}$ \\
$76$  & $ \{u^{\mu},[u^{\nu},h^{\lambda\rho}]\}\gamma_{5}\gamma_{\mu}D_{\nu\lambda\rho}  +h.c. $   & $-24im^{3}  \{S \cdot u ,[\upsilon \cdot u ,h^{\lambda\rho}]\}\upsilon_{\lambda}\upsilon_{\rho} $ \\
$77$  & $\{u^{\nu},[u^{\mu},h^{\lambda\rho}]\}\gamma_{5}\gamma_{\mu}D_{\nu\lambda\rho}  +h.c.  $  & $-24im^{3} \{\upsilon \cdot u ,[S \cdot u ,h^{\lambda\rho}]\}\upsilon_{\lambda}\upsilon_{\rho}  $ \\
$78$  & $  \left \langle [u_{\mu},u_{\nu}]h_{\lambda\rho} \right \rangle  \gamma_{5}\gamma^{\mu}D^{\nu\lambda\rho} +h.c. $   & $-24im^{3} \left \langle [S \cdot u,\upsilon \cdot u ]h_{\lambda\rho} \right \rangle \upsilon_{\lambda}\upsilon_{\rho} $ \\
$79$  & $i\{[u_{\mu},u_{\nu}],f_{+}^{\mu\nu}\}$   & $i\{[u_{\mu},u_{\nu}],f_{+}^{\mu\nu}\}$ \\
$80$  & $ i\{u_{\mu},[u_{\nu},f_{+}^{\mu\nu}]\}$   & $i\{u_{\mu},[u_{\nu},f_{+}^{\mu\nu}]\}$ \\
$81$  & $i \left \langle [u_{\mu},u_{\nu}]f_{+}^{\mu\nu} \right \rangle $   & $ i \left \langle [u_{\mu},u_{\nu}]f_{+}^{\mu\nu} \right \rangle  $ \\
$82$  & $\{u^{\mu},\{u_{\mu},f_{+}^{\nu\lambda}\}\}\sigma_{\nu\lambda} $   & $ -2i\{u^{\mu},\{u_{\mu},f_{+}^{\nu\lambda}\}\}[S_{\nu},S_{\lambda}]  $  \\
$83$  & $[u^{\mu},[u_{\mu},f_{+}^{\nu\lambda}]]\sigma_{\nu\lambda}  $   & $-2i[u^{\mu},[u_{\mu},f_{+}^{\nu\lambda}]][S_{\nu},S_{\lambda}] $ \\
$84$  & $\{u_{\mu},\{u^{\lambda},f_{+}^{\mu\nu}\}\}\sigma_{\nu\lambda} $  & $-2i\{u_{\mu},\{u^{\lambda},f_{+}^{\mu\nu}\}\} [S_{\nu},S_{\lambda}] $  \\
$85$  & $[u_{\mu},[u^{\lambda},f_{+}^{\mu\nu}]]\sigma_{\nu\lambda} $   & $-2i[u_{\mu},[u^{\lambda},f_{+}^{\mu\nu}]] [S_{\nu},S_{\lambda}] $ \\
$86$  & $\{u^{\lambda},\{u_{\mu},f_{+}^{\mu\nu}\}\}\sigma_{\nu\lambda} $   & $-2i\{u^{\lambda},\{u_{\mu},f_{+}^{\mu\nu}\}\} [S_{\nu},S_{\lambda}] $ \\
$87$  & $f_{+}^{\mu\nu} \left \langle u^{2} \right \rangle \sigma_{\mu\nu} $   & $-2i[S_{\mu},S_{\nu}]f_{+}^{\mu\nu} \left \langle u^{2} \right \rangle  $ \\
$88$  & $u^{\mu} \left \langle f_{+\nu\lambda}u_{\mu} \right \rangle \sigma^{\nu\lambda}$   & $ -2iu^{\mu}[S^{\nu},S^{\lambda}]\left \langle f_{+\nu\lambda}u_{\mu} \right \rangle $ \\
$89$  & $f_{+}^{\mu\nu} \left \langle u_{\nu}u_{\lambda} \right \rangle {\sigma_{\mu}}^{\lambda} $   & $ -2if_{+}^{\mu\nu}[S_{\mu},S^{\lambda}] \left \langle u_{\nu}u_{\lambda} \right \rangle $  \\
$90$  & $ u^{\mu} \left \langle {f_{+\nu}}^{\lambda}u_{\lambda} \right \rangle {\sigma_{\mu}}^{\nu}$   & $-2iu^{\mu}[S_{\mu},S^{\nu}] \left \langle {f_{+\nu}}^{\lambda}u_{\lambda} \right \rangle  $ \\ 
$91$  & $ u^{\mu} \left \langle f_{+\mu\nu}u_{\lambda} \right \rangle \sigma^{\nu\lambda} $   & $ -2iu^{\mu}[S^{\nu},S^{\lambda}]\left \langle f_{+\mu\nu}u_{\lambda} \right \rangle $ \\
$92$  & $i\{[u_{\mu},u^{\lambda}],f_{+}^{\mu\nu}\} D_{\nu\lambda} +h.c. $   & $ -4im^{2}\{[u_{\mu},\upsilon \cdot u],f_{+}^{\mu\nu}\} \upsilon_{\nu}$ \\
$93$  & $ i[\{u_{\mu},u^{\lambda}\},f_{+}^{\mu\nu}] D_{\nu\lambda}+h.c.$   &  $ -4im^{2}[\{u_{\mu},\upsilon \cdot u \},f_{+}^{\mu\nu}] \upsilon_{\nu} $ \\
$94$  & $i\{u_{\mu},[u^{\lambda},f_{+}^{\mu\nu}]\}D_{\nu\lambda}+h.c. $  & $ -4im^{2}\{u_{\mu},[\upsilon \cdot u ,f_{+}^{\mu\nu}]\}\upsilon_{\nu}$ \\
$95$  & $ i \left \langle [u_{\nu},u_{\lambda}]{f_{+\mu}}^{\lambda} \right \rangle D^{\mu\nu}+h.c.$   & $-4im^{2} \left \langle [\upsilon \cdot u,u_{\lambda}]{f_{+\mu}}^{\lambda} \right \rangle \upsilon^{\mu}  $ \\
$96$  & $ \{u^{\lambda},\{u^{\rho},f_{+}^{\mu\nu}\}\}\sigma_{\mu\nu}D_{\lambda\rho} +h.c.  $   & $8im^{2} \{\upsilon \cdot u,\{\upsilon \cdot u ,f_{+}^{\mu\nu}\}\}[S_{\mu},S_{\nu}] $\\
$97$  & $[u^{\lambda},[u^{\rho},f_{+}^{\mu\nu}]]\sigma_{\mu\nu}D_{\lambda\rho}+h.c.  $   & $8im^{2}[\upsilon \cdot u ,[\upsilon \cdot u,f_{+}^{\mu\nu}]] [S_{\mu}, S_{\nu}] $ \\
$98$  & $\{u^{\lambda},\{u^{\rho},f_{+}^{\mu\nu}\}\}\sigma_{\mu\lambda}D_{\nu\rho} +h.c. $   & $8im^{2}\{u^{\lambda},\{\upsilon \cdot u, f_{+}^{\mu\nu}\}\}[S_{\mu},S_{\lambda}]\upsilon_{\nu} $ \\
$99$  & $[u^{\lambda},[u^{\rho},f_{+}^{\mu\nu}]]\sigma_{\mu\lambda}D_{\nu\rho} +h.c. $   & $8im^{2} [u^{\lambda},[\upsilon \cdot u ,f_{+}^{\mu\nu}]][S_{\mu},S_{\lambda}]\upsilon_{\nu}  $ \\
$100$  & $\{u^{\rho},\{u^{\lambda},f_{+}^{\mu\nu}\}\}\sigma_{\mu\lambda}D_{\nu\rho} +h.c. $   & $8im^{2}\{\upsilon \cdot u ,\{u^{\lambda},f_{+}^{\mu\nu}\}\}[S_{\mu},S_{\lambda}]\upsilon_{\nu}$ \\
$101$  & $ f_{+}^{\mu\nu} \left \langle u_{\lambda}u_{\rho} \right \rangle \sigma_{\mu\nu}D^{\lambda\rho}+h.c.$   & $8im^{2} f_{+}^{\mu\nu}[S_{\mu},S_{\nu}] \left \langle (\upsilon \cdot u)^{2} \right \rangle $ \\
$102$  & $ f_{+}^{\mu\nu} \left \langle u_{\lambda}u_{\rho} \right \rangle {\sigma_{\mu}}^{\lambda}{D_{\nu}}^{\rho}+h.c.$  & $8im^{2}f_{+}^{\mu\nu}[S_{\mu},S^{\lambda}]\upsilon_{\nu}\left \langle u_{\lambda}\upsilon \cdot u \right \rangle  $  \\
$103$  & $u^{\mu} \left \langle f_{+\nu\lambda}u_{\rho} \right \rangle {\sigma_{\mu}}^{\nu}D^{\lambda\rho} +h.c.$   & $8im^{2}u^{\mu}[S_{\mu},S^{\nu}] \left \langle f_{+\nu\lambda}\upsilon \cdot u \right \rangle \upsilon^{\lambda} $ \\
$104$  & $u^{\mu} \left \langle f_{+\nu\lambda}u_{\rho} \right \rangle \sigma^{\nu\lambda}{D_{\mu}}^{\rho} +h.c.$   & $8im^{2}\upsilon \cdot u [S^{\nu},S^{\lambda}]  \left \langle f_{+\nu\lambda}\upsilon \cdot u \right \rangle $ \\
$105$  & $ u^{\mu} \left \langle f_{+\nu\rho}u_{\lambda} \right \rangle \sigma^{\nu\lambda}{D_{\mu}}^{\rho}+h.c. $   & $8im^{2}\upsilon \cdot u[S^{\nu},S^{\lambda}]  \left \langle f_{+\nu\rho}u_{\lambda} \right \rangle \upsilon^{\rho}$ \\
$106$  & $ i \{[u^{\mu},u^{\nu}],\widetilde{\chi}_{-}\} \gamma_{5}\gamma_{\mu}D_{\nu}  +h.c.$   & $ -4m \{[S \cdot u,\upsilon \cdot u ],\widetilde{\chi}_{-}\} $ \\
$107$  & $i [\{u^{\mu},u^{\nu}\},\widetilde{\chi}_{-} ]\gamma_{5}\gamma_{\mu}D_{\nu}  +h.c. $   & $-4m [\{S \cdot u,\upsilon \cdot u\},\widetilde{\chi}_{-} ]$ \\
$108$  & $ i \{u^{\mu},[u^{\nu},\widetilde{\chi}_{-}]\} \gamma_{5}\gamma_{\mu}D_{\nu}+h.c.$  & $-4m\{S \cdot u,[\upsilon \cdot u, \widetilde{\chi}_{-}]\}$ \\
$109$  & $ i \left \langle  [u_{\mu},u_{\nu}]\widetilde{\chi}_{-} \right \rangle \gamma_{5}\gamma^{\mu}D^{\nu} +h.c. $  & $-4m\left \langle  [S \cdot u ,\upsilon \cdot u ]\widetilde{\chi}_{-} \right \rangle $\\
$110$  & $i [u_{\mu},u_{\nu}] \left \langle  \chi_{-} \right \rangle \gamma_{5}\gamma^{\mu}D^{\nu}+h.c. $ & $-4m [S \cdot u,\upsilon \cdot u ] \left \langle  \chi_{-} \right \rangle $\\
$111$  & $\{u_{\mu},\{u^{\mu},\widetilde{\chi}_{+}\}\}$ & $ \{u_{\mu},\{u^{\mu},\widetilde{\chi}_{+}\}\}$\\
$112$  & $[u_{\mu},[u^{\mu},\widetilde{\chi}_{+}]] $ & $ [u_{\mu},[u^{\mu},\widetilde{\chi}_{+}]]$ \\
$113$  & $ u^{\mu} \left \langle   u_{\mu}\widetilde{\chi}_{+}     \right \rangle$  & $u^{\mu}\left \langle   u_{\mu}\widetilde{\chi}_{+}     \right \rangle $\\
$114$  & $\widetilde{\chi}_{+} \left \langle   u^{2}    \right \rangle $ & $\widetilde{\chi}_{+} \left \langle   u^{2}    \right \rangle$\\
$115$  & $ u^{2} \left \langle   \chi_{+}     \right \rangle$ & $ u^{2} \left \langle   \chi_{+}     \right \rangle $ \\
$116$  & $i\{u^{\mu},[u^{\nu},\widetilde{\chi}_{+}]\}\sigma_{\mu\nu}$  & $2\{u^{\mu},[u^{\nu},\widetilde{\chi}_{+}]\}[S_{\mu},S_{\nu}]$ \\
$117$  & $i[u^{\mu},\{u^{\nu},\widetilde{\chi}_{+}\}]\sigma_{\mu\nu}$  & $2[u^{\mu},\{u^{\nu},\widetilde{\chi}_{+}\}][S_{\mu},S_{\nu}]$ \\
$118$  & $i \left \langle   [u_{\mu},u_{\nu}]\widetilde{\chi}_{+}     \right \rangle  \sigma^{\mu\nu}$  & $2[S^{\mu},S^{\nu}] \left \langle   [u_{\mu},u_{\nu}]\widetilde{\chi}_{+}     \right \rangle $\\
$119$  & $i[u^{\mu},u^{\nu}]\left \langle  \chi_{+}  \right \rangle  \sigma_{\mu\nu} $ & $2[u^{\mu},u^{\nu}][S_{\mu},S_{\nu}]\left \langle  \chi_{+}  \right \rangle $  \\
$120$  & $\{u^{\mu},\{u^{\nu},\widetilde{\chi}_{+}\}\}D_{\mu\nu} +h.c.  $ & $-4m^{2}\{\upsilon \cdot u,\{\upsilon \cdot u,\widetilde{\chi}_{+}\}\} $\\
$121$  & $[u^{\mu},[u^{\nu},\widetilde{\chi}_{+}]]D_{\mu\nu} +h.c. $ & $-4m^{2}[\upsilon \cdot u,[\upsilon \cdot u,\widetilde{\chi}_{+}]]$\\
$122$  & $u^{\mu}  \left \langle   u_{\nu}\widetilde{\chi}_{+}     \right \rangle {D_{\mu}}^{\nu} +h.c.  $ & $-4m^{2}\upsilon \cdot u  \left \langle  \upsilon \cdot  u \widetilde{\chi}_{+}     \right \rangle $ \\
$123$  & $\widetilde{\chi}_{+} \left \langle    u_{\mu}u_{\nu}   \right \rangle D^{\mu\nu}+h.c.$ & $-4m^{2}\widetilde{\chi}_{+}  \left \langle   (\upsilon \cdot  u)^{2} \right \rangle $  \\
$124$  & $\{u^{\mu},u^{\nu}\}  \left \langle   \chi_{+}     \right \rangle D_{\mu\nu} +h.c. $ & $-8m^{2}(\upsilon \cdot u)^{2} \left \langle   \chi_{+}     \right \rangle $ \\
$125$  & $[u^{\mu},[D^{\nu}, \widetilde{\chi}_{+}]]\gamma_{5}\gamma_{\nu}D_{\mu}+h.c.  $ & $4im[\upsilon \cdot u ,[S \cdot D, \widetilde{\chi}_{+}]] $\\
$126$  & $[u^{\mu},[D^{\nu}, \widetilde{\chi}_{+}]]\gamma_{5}\gamma_{\mu}D_{\nu}+h.c.  $ & $4im[S \cdot u,[\upsilon \cdot D, \widetilde{\chi}_{+}]]$\\
$127$  & $i\{u^{\mu},[D_{\mu},\widetilde{\chi}_{-}]\}   $ & $i\{u^{\mu},[D_{\mu},\widetilde{\chi}_{-}]\}  $  \\
$128$  & $ i  \left \langle u^{\mu}[D_{\mu},\widetilde{\chi}_{-}] \right \rangle$ & $i \left \langle u^{\mu}[D_{\mu},\widetilde{\chi}_{-}] \right \rangle$\\
$129$  & $i u^{\mu} \left \langle [D_{\mu},\chi_{-}] \right \rangle$ & $ iu^{\mu}  \left \langle [D_{\mu},\chi_{-}] \right \rangle$ \\
$130$  & $ [u^{\mu},[D^{\nu},\widetilde{\chi}_{-}]]\sigma_{\mu\nu}$ & $-2i[u^{\mu},[D^{\nu},\widetilde{\chi}_{-}]][S_{\mu},S_{\nu}]$ \\
$131$  & $i\{u^{\mu},[D^{\nu},\widetilde{\chi}_{-}]\}D_{\mu\nu}+h.c. $ & $ -4im^{2}\{\upsilon \cdot u ,[\upsilon \cdot D,\widetilde{\chi}_{-}]\}$ \\
$132$  & $ i \left \langle u^{\mu}[D^{\nu},\widetilde{\chi}_{-}]\right \rangle D_{\mu\nu}+h.c.$ & $ -4im^{2}\left \langle \upsilon \cdot u [\upsilon \cdot D ,\widetilde{\chi}_{-}]\right \rangle $\\
$133$  & $iu^{\mu} \left \langle [D_{\nu},\chi_{-}] \right \rangle {D_{\mu}}^{\nu}+h.c. $ & $-4im^{2} \upsilon \cdot u \left \langle [\upsilon \cdot D ,\chi_{-}] \right \rangle  $ \\
$134$  & $i\{[D^{\mu},{f_{+\mu}}^{\nu}],u^{\lambda}\} \gamma_{5}\gamma_{\nu}D_{\lambda} +h.c. $ & $-4m\{[D^{\mu},{f_{+\mu}}^{\nu}],\upsilon \cdot u\} S_{\nu}$ \\
$135$  & $ i\{[D^{\mu},{f_{+\mu}}^{\nu}],u^{\lambda}\} \gamma_{5}\gamma_{\lambda}D_{\nu}+h.c.$ & $-4m\{[D^{\mu},{f_{+\mu}}^{\nu}],S \cdot u \} \upsilon_{\nu}$ \\
$136$  & $ i\{[D^{\mu},{f_{+\lambda}}^{\nu}],u^{\lambda} \}\gamma_{5}\gamma_{\mu}D_{\nu} +h.c.$ & $-4m\{[S \cdot D,{f_{+\lambda}}^{\nu}],u^{\lambda} \}\upsilon_{\nu}$ \\
$137$  & $ i\{[D^{\mu},f_{+}^{\nu\lambda}],u_{\mu} \}\gamma_{5}\gamma_{\nu}D_{\lambda} +h.c.$ & $ -4m\{[D^{\mu},f_{+}^{\nu\lambda}],u_{\mu} \}S_{\nu}\upsilon_{\lambda}$\\
$138$  & $ i \left \langle [D_{\mu},{f_{+\nu}}^{\lambda}]u_{\lambda} \right \rangle \gamma_{5}\gamma^{\mu}D^{\nu}+h.c.$  & $ -4m \left \langle [S \cdot D ,{f_{+\nu}}^{\lambda}]u_{\lambda} \right \rangle \upsilon_{\nu}$\\
$139$  & $i \left \langle [D^{\lambda},f_{+\mu\lambda}]u_{\nu} \right \rangle \gamma_{5}\gamma^{\mu}D^{\nu}  +h.c. $ & $-4mS^{\mu} \left \langle [D^{\lambda},f_{+\mu\lambda}]\upsilon \cdot u \right \rangle $\\
$140$  & $ i \left \langle [D^{\mu},f_{+}^{\nu\lambda}]u_{\mu} \right \rangle \gamma_{5}\gamma_{\nu}D_{\lambda}+h.c.$ & $ -4m \left \langle [D^{\mu},f_{+}^{\nu\lambda}]u_{\mu} \right \rangle S_{\nu}\upsilon_{\lambda}$  \\
$141$  & $ i \left \langle [D^{\lambda},f_{+\nu\lambda}]u_{\mu} \right \rangle \gamma_{5}\gamma^{\mu}D^{\nu}  +h.c.$ & $ -4m\left \langle [D^{\lambda},f_{+\nu\lambda}]S \cdot u \right \rangle \upsilon^{\nu}$ \\
$142$  & $\{u^{\mu},[D^{\nu},f_{-\mu\nu}] \}$ & $ \{u^{\mu},[D^{\nu},f_{-\mu\nu}] \} $ \\
$143$  & $ \left \langle u^{\mu}[D^{\nu},f_{-\mu\nu}] \right \rangle   $ & $ \left \langle u^{\mu}[D^{\nu},f_{-\mu\nu}] \right \rangle $\\
$144$  & $i[u^{\mu},[D_{\mu},f^{\nu\lambda}_{-}]]\sigma_{\nu\lambda}$  & $2[u^{\mu},[D_{\mu},f^{\nu\lambda}_{-}]][S_{\nu},S_{\lambda}] $\\
$145$  & $ i[u^{\mu},[D^{\nu},{f_{-\nu}}^{\lambda}]]\sigma_{\mu\lambda}$ & $ 2[u^{\mu},[D^{\nu},{f_{-\nu}}^{\lambda}]][S_{\mu},S_{\lambda}]$\\
$146$  & $i[u_{\nu},[D^{\mu},f^{\nu\lambda}_{-}]]\sigma_{\mu\lambda}$ & $2[u_{\nu},[D^{\mu},f^{\nu\lambda}_{-}]][S_{\mu},S_{\lambda}] $\\
$147$  & $\{u^{\lambda},[D_{\mu},f_{-}^{\mu\nu}]\}D_{\nu\lambda}  +h.c.  $  & $-4m^{2}\{\upsilon \cdot u ,[D_{\mu},f_{-}^{\mu\nu}]\}\upsilon_{\nu}$\\
$148$  & $\{u_{\mu},[D^{\lambda},f_{-}^{\mu\nu}]\}D_{\nu\lambda}  +h.c.$ & $-4m^{2}\{u_{\mu},[\upsilon \cdot D,f_{-}^{\mu\nu}]\}\upsilon_{\nu}$ \\
$149$  & $  \left \langle u_{\mu}[D^{\lambda},f_{-}^{\mu\nu} ]\right \rangle  D_{\nu\lambda} +h.c.$ & $ -4m^{2}\upsilon_{\nu} \left \langle u_{\mu}[\upsilon \cdot D ,f_{-}^{\mu\nu} ]\right \rangle $\\
$150$  & $ \left \langle u^{\lambda}[D_{\mu},f_{-}^{\mu\nu}] \right \rangle D_{\nu\lambda}+h.c.  $ & $-4m^{2}\upsilon_{\nu}\left \langle \upsilon \cdot u [D_{\mu},f_{-}^{\mu\nu}] \right \rangle$\\
$151$  & $\widetilde{\chi}_{+}^{2} $ & $ \widetilde{\chi}_{+}^{2} $ \\
$152$  & $  \left \langle  \widetilde{\chi}_{+}^{2} \right \rangle $ & $ \left \langle  \widetilde{\chi}_{+}^{2} \right \rangle $\\
$153$  & $\widetilde{\chi}_{+}   \left \langle  \chi_{+} \right \rangle  $  & $ \widetilde{\chi}_{+}  \left \langle  \chi_{+} \right \rangle  $ \\
$154$  & $  \left \langle  \chi_{+} \right \rangle ^{2} $  & $  \left \langle  \chi_{+} \right \rangle^{2}  $ \\
$155$  & $\{f_{+}^{\mu\nu},\widetilde{ \chi}_{+} \}\sigma_{\mu\nu}  $ & $-2i\{f_{+}^{\mu\nu},\widetilde{ \chi}_{+} \}[S_{\mu},S_{\nu}] $\\
$156$  & $ \left \langle  f_{+\mu\nu}\widetilde{\chi}_{+} \right \rangle \sigma^{\mu\nu}   $ & $-2i[S^{\mu},S^{\nu}] \left \langle  f_{+\mu\nu}\widetilde{\chi}_{+} \right \rangle $ \\
$157$  & $f_{+}^{\mu\nu}  \left \langle \chi_{+} \right \rangle \sigma_{\mu\nu}  $  & $-2if_{+}^{\mu\nu}[S_{\mu},S_{\nu}]  \left \langle \chi_{+} \right \rangle $\\
$158$  & $[\widetilde{\chi}_{+},h^{\mu\nu}]\gamma_{5}\gamma_{\mu}D_{\nu}+h.c.  $  & $4im[\widetilde{\chi}_{+},h^{\mu\nu}]S_{\mu}\upsilon_{\nu}  $\\
$159$  & $[\widetilde{\chi}_{+},f_{-}^{\mu\nu}]\gamma_{5}\gamma_{\mu}D_{\nu} +h.c. $  & $4im[\widetilde{\chi}_{+},f_{-}^{\mu\nu}]S_{\mu}\upsilon_{\nu} $  \\
$160$  & $[D^{2},\widetilde{\chi}_{+}]  $ & $[D^{2},\widetilde{\chi}_{+}]$\\
$161$  & $ \left \langle  [D^{2},\chi_{+}]\right \rangle $ & $  \left \langle  [D^{2},\chi_{+}]\right \rangle  $\\
$162$  & $\widetilde{\chi}_{-}^{2}    $  & $\widetilde{\chi}_{-}^{2}  $  \\
$163$  & $\left \langle \widetilde{\chi}_{-}^{2} \right \rangle$ & $ \left \langle \widetilde{\chi}_{-}^{2} \right \rangle$ \\
$164$  & $  \widetilde{\chi}_{-} \left \langle \chi_{-} \right \rangle$ & $  \widetilde{\chi}_{-}\left \langle \chi_{-} \right \rangle $\\
$165$  & $ \left \langle \chi_{-} \right \rangle^{2}$ & $ \left \langle \chi_{-} \right \rangle^{2} $\\
$166$  & $\{f_{+}^{\mu\nu} ,\widetilde{\chi}_{-}\}\gamma_{5}\gamma_{\mu}D_{\nu} +h.c.$  & $4im\{f_{+}^{\mu\nu} ,\widetilde{\chi}_{-}\}S_{\mu}\upsilon_{\nu} $ \\
$167$  & $ \left \langle f_{+}^{\mu\nu}\widetilde{\chi}_{-} \right \rangle \gamma_{5}\gamma_{\mu}D_{\nu}+h.c. $  & $4im \left \langle f_{+}^{\mu\nu}\widetilde{\chi}_{-} \right \rangle S_{\mu}\upsilon_{\nu}  $\\
$168$  & $f_{+}^{\mu\nu} \left \langle \chi_{-} \right \rangle \gamma_{5}\gamma_{\mu}D_{\nu} +h.c. $  & $4im f_{+}^{\mu\nu}\left \langle \chi_{-} \right \rangle S_{\mu}\upsilon_{\nu}  $\\
$169$  & $i\{h^{\mu\nu},\widetilde{\chi}_{-}\}D_{\mu\nu}+h.c. $ & $-4im^{2}\{h^{\mu\nu},\widetilde{\chi}_{-}\}\upsilon_{\mu}\upsilon_{\nu}  $  \\
$170$  & $i \left \langle h_{\mu\nu}\widetilde{\chi}_{-} \right \rangle D^{\mu\nu}+h.c.$ & $-4im^{2}\upsilon^{\mu}\upsilon^{\nu}\left \langle h_{\mu\nu}\widetilde{\chi}_{-} \right \rangle  $ \\
$171$  & $ ih_{\mu\nu} \left \langle \chi_{-} \right \rangle D^{\mu\nu} +h.c. $ & $-4im^{2}h_{\mu\nu}\upsilon^{\mu}\upsilon^{\nu}\left \langle \chi_{-} \right \rangle  $\\
$172$  & $[f_{-}^{\mu\nu},\widetilde{\chi}_{-}]\sigma_{\mu\nu} $ & $ -2i[f_{-}^{\mu\nu},\widetilde{\chi}_{-}][S_{\mu},S_{\nu}] $ \\
$173$  & $  \{f_{+}^{\mu\nu},f_{+\mu\nu}\} $ & $  \{f_{+}^{\mu\nu},f_{+\mu\nu}\}$ \\
$174$  & $  \left \langle f_{+}^{\mu\nu}f_{+\mu\nu} \right \rangle $  & $  \left \langle f_{+}^{\mu\nu}f_{+\mu\nu} \right \rangle $ \\
$175$  & $ i[f_{+}^{\mu\nu},{f_{+\mu}}^{\lambda}]\sigma_{\nu\lambda} $ & $ 2[f_{+}^{\mu\nu},{f_{+\mu}}^{\lambda}][S_{\nu},S_{\lambda}]$\\
$176$  & $\{f_{+}^{\mu\nu},{f_{+\mu}}^{\lambda}\}D_{\nu\lambda} +h.c.  $ & $-4m^{2}\{f_{+}^{\mu\nu},{f_{+\mu}}^{\lambda}\}\upsilon_{\nu}\upsilon_{\lambda} $  \\
$177$  & $ \left \langle {f_{+\mu}}^{\lambda}f_{+\nu\lambda} \right \rangle D^{\mu\nu} +h.c.$  & $-4m^{2}\upsilon^{\mu}\upsilon^{\nu} \left \langle {f_{+\mu}}^{\lambda}f_{+\nu\lambda} \right \rangle $\\
$178$  & $i[f_{+}^{\mu\nu},f_{+}^{\lambda\rho}]\sigma_{\mu\lambda}D_{\nu\rho} +h.c. $ & $-8m^{2}[f_{+}^{\mu\nu},f_{+}^{\lambda\rho}][S_{\mu},S_{\lambda}]\upsilon_{\nu}\upsilon_{\rho} $ \\
$179$  & $ i\{f_{+}^{\mu\nu},{h_{\mu}}^{\lambda}\}\gamma_{5}\gamma_{\nu}D_{\lambda} +h.c.  $ & $-4m \{f_{+}^{\mu\nu},{h_{\mu}}^{\lambda}\}S_{\nu}\upsilon_{\lambda}$ \\
$180$  & $ i\{f_{+}^{\mu\nu},{h_{\mu}}^{\lambda}\}\gamma_{5}\gamma_{\lambda}D_{\nu} +h.c.$ & $ -4m\{f_{+}^{\mu\nu},{h_{\mu}}^{\lambda}\}S_{\lambda}\upsilon_{\nu} $\\
$181$  & $i \left \langle {f_{+\mu}}^{\lambda}h_{\nu\lambda} \right \rangle \gamma_{5}\gamma^{\mu}D^{\nu} +h.c. $ & $-4mS^{\mu}\upsilon^{\nu} \left \langle {f_{+\mu}}^{\lambda}h_{\nu\lambda} \right \rangle  $\\
$182$  & $ i\left \langle {f_{+\nu}}^{\lambda}h_{\mu\lambda} \right \rangle \gamma_{5}\gamma^{\mu}D^{\nu} +h.c.$ & $ -4mS^{\mu}\upsilon^{\nu} \left \langle {f_{+\nu}}^{\lambda}h_{\mu\lambda} \right \rangle $\\
$183$  & $i\epsilon^{\mu\nu\lambda\rho}[f_{+\mu\nu},{h_{\lambda}}^{\sigma}]D_{\rho\sigma}+h.c. $& $-4im^{2}\epsilon^{\mu\nu\lambda\rho}[f_{+\mu\nu},{h_{\lambda}}^{\sigma}]\upsilon_{\rho}\upsilon_{\sigma}$ \\
$184$  & $i\{f_{+}^{\mu\nu},h^{\lambda\rho}\}\gamma_{5}\gamma_{\mu}D_{\nu\lambda\rho} +h.c.  $& $24m^{3}\{f_{+}^{\mu\nu},h^{\lambda\rho}\}S_{\mu}\upsilon_{\nu}\upsilon_{\lambda}\upsilon_{\rho} $ \\
$185$  & $i \left \langle f_{+\mu\nu}h_{\lambda\rho} \right \rangle \gamma_{5}\gamma^{\mu}D^{\nu\lambda\rho}+h.c. $ & $24m^{3}S^{\mu}\upsilon^{\nu}\upsilon^{\lambda}\upsilon^{\rho} \left \langle f_{+\mu\nu}h_{\lambda\rho} \right \rangle   $\\
$186$  & $i\epsilon^{\mu\nu\lambda\rho}[f_{+\mu\nu},f_{-\lambda\rho}]$& $i\epsilon^{\mu\nu\lambda\rho}[f_{+\mu\nu},f_{-\lambda\rho}] $ \\
$187$  & $ i\{f_{+}^{\mu\nu},{f_{-\mu}}^{\lambda}\}\gamma_{5}\gamma_{\nu}D_{\lambda} +h.c. $ & $ -4mS_{\nu}\upsilon_{\lambda}\{f_{+}^{\mu\nu},{f_{-\mu}}^{\lambda}\}$\\
$188$  & $i\{f_{+}^{\mu\nu},{f_{-\mu}}^{\lambda}\}\gamma_{5}\gamma_{\lambda}D_{\nu} +h.c.$  & $-4mS_{\lambda}\upsilon_{\nu}\{f_{+}^{\mu\nu},{f_{-\mu}}^{\lambda}\}$\\
$189$  & $i \left \langle {f_{+\mu}}^{\lambda}f_{-\nu\lambda} \right \rangle \gamma_{5}\gamma^{\mu}D^{\nu} +h.c. $ & $-4mS^{\mu}\upsilon^{\nu} \left \langle {f_{+\mu}}^{\lambda}f_{-\nu\lambda} \right \rangle$\\
$190$  & $i \left \langle {f_{+\nu}}^{\lambda}f_{-\mu\lambda} \right \rangle \gamma_{5}\gamma^{\mu}D^{\nu}+h.c.$ & $-4mS^{\mu}\upsilon^{\nu} \left \langle {f_{+\nu}}^{\lambda}f_{-\mu\lambda} \right \rangle $  \\
$191$  & $i\epsilon^{\mu\nu\lambda\rho}[f_{+\mu\nu},{f_{-\lambda}}^{\sigma}]D_{\rho\sigma}+h.c. $ & $-4im^{2}\epsilon^{\mu\nu\lambda\rho}[f_{+\mu\nu},{f_{-\lambda}}^{\sigma}]\upsilon_{\rho}\upsilon_{\sigma}$\\
$192$  & $i\epsilon^{\mu\nu\lambda\rho}[{f_{+\mu}}^{\sigma},f_{-\nu\lambda}]D_{\rho\sigma} +h.c. $ & $-4im^{2}\epsilon^{\mu\nu\lambda\rho}[{f_{+\mu}}^{\sigma},f_{-\nu\lambda}]\upsilon_{\rho}\upsilon_{\sigma}$\\
$193$  & $[D^{2},f_{+}^{\nu\lambda}]\sigma_{\nu\lambda}  $ & $-2i[D^{2},f_{+}^{\nu\lambda}][S_{\nu},S_{\lambda}]$\\
$194$  & $[{D_{\mu}}^{\nu},f_{+\nu\lambda}]\sigma^{\mu\lambda}  $ & $-2i[{D_{\mu}}^{\nu},f_{+\nu\lambda}][S^{\mu},S^{\lambda}]$\\
$195$  & $[D^{\mu\nu},f_{+}^{\lambda\rho}]\sigma_{\lambda\rho}D_{\mu\nu}  +h.c. $ & $8im^{2}[D^{\mu\nu},f_{+}^{\lambda\rho}][S_{\lambda},S_{\rho}]\upsilon_{\mu}\upsilon_{\nu}$\\
$196$  & $[D^{\mu\nu},f_{+}^{\lambda\rho}]\sigma_{\mu\lambda}D_{\nu\rho}  +h.c. $ & $8im^{2}[D^{\mu\nu},f_{+}^{\lambda\rho}][S_{\mu},S_{\lambda}]\upsilon_{\nu}\upsilon_{\rho}$\\
$197$  & $\{h^{\mu\nu},h_{\mu\nu}\} $ & $ \{h^{\mu\nu},h_{\mu\nu}\}$ \\
$198$  & $ \left \langle  h^{\mu\nu}h_{\mu\nu}  \right \rangle  $ & $\left \langle  h^{\mu\nu}h_{\mu\nu}  \right \rangle $\\
$199$  & $i [h^{\mu\nu},{h_{\mu}}^{\lambda}]\sigma_{\nu\lambda} $ & $2 [h^{\mu\nu},{h_{\mu}}^{\lambda}][S_{\nu},S_{\lambda}]$ \\ 
$200$  & $\{h^{\mu\nu},{h_{\mu}}^{\lambda}\}D_{\nu\lambda} +h.c. $& $-4m^{2}\{h^{\mu\nu},{h_{\mu}}^{\lambda}\}\upsilon_{\nu}\upsilon_{\lambda} $ \\
$201$  & $ \left \langle  {h_{\mu}}^{\lambda}h_{\nu\lambda}  \right \rangle D^{\mu\nu} +h.c. $ & $-4m^{2}\upsilon^{\mu}\upsilon^{\nu}\left \langle  {h_{\mu}}^{\lambda}h_{\nu\lambda}  \right \rangle $\\
$202$  & $i[h^{\mu\nu},h^{\lambda\rho}]\sigma_{\mu\lambda}D_{\nu\rho} +h.c. $ & $-8m^{2}[h^{\mu\nu},h^{\lambda\rho}][S_{\mu},S_{\lambda}]\upsilon_{\nu}\upsilon_{\rho}$  \\
$203$  & $ \{h^{\mu\nu},h^{\lambda\rho} \}D_{\mu\nu\lambda\rho}  +h.c.$ & $ 48m^{4}\{h^{\mu\nu},h^{\lambda\rho} \}\upsilon_{\mu}\upsilon_{\nu}\upsilon_{\lambda}\upsilon_{\rho}$\\
$204$  & $ \left \langle h_{\mu\nu}h_{\lambda\rho}  \right \rangle D^{\mu\nu\lambda\rho} +h.c.$ & $48m^{4}\upsilon^{\mu}\upsilon^{\nu}\upsilon^{\lambda}\upsilon^{\rho} \left \langle h_{\mu\nu}h_{\lambda\rho}  \right \rangle $ \\
$205$  & $i [f_{-}^{\mu\nu},{h_{\mu}}^{\lambda}]\sigma_{\nu\lambda} $ & $2 [f_{-}^{\mu\nu},{h_{\mu}}^{\lambda}][S_{\nu},S_{\lambda}] $\\
$206$  & $ \{f^{\mu\nu}_{-},{h_{\mu}}^{\lambda}\}D_{\nu\lambda} + h.c.$ & $-4m^{2}\upsilon_{\nu}\upsilon_{\lambda}\{f^{\mu\nu}_{-},{h_{\mu}}^{\lambda}\}$  \\
$207$  & $ \left \langle  {f_{-\mu}}^{\lambda}h_{\nu\lambda}  \right \rangle D^{\mu\nu} +h.c.$ & $ -4m^{2}\upsilon^{\mu}\upsilon^{\nu}\left \langle  {f_{-\mu}}^{\lambda}h_{\nu\lambda}  \right \rangle$ \\
$208$  & $i[f^{\mu\nu}_{-},h^{\lambda\rho}]\sigma_{\mu\nu}D_{\lambda\rho}+ h.c. $ & $-8m^{2}[f^{\mu\nu}_{-},h^{\lambda\rho}][S_{\mu},S_{\nu}]\upsilon_{\lambda}\upsilon_{\rho} $ \\
$209$  & $i[f^{\mu\nu}_{-},h^{\lambda\rho}]\sigma_{\mu\lambda}D_{\nu\rho} + h.c. $ & $-8m^{2}[f^{\mu\nu}_{-},h^{\lambda\rho}][S_{\mu},S_{\lambda}]\upsilon_{\nu}\upsilon_{\rho} $\\
$210$  & $\{f_{-}^{\mu\nu},f_{-\mu\nu}\} $ & $\{f_{-}^{\mu\nu},f_{-\mu\nu}\}$ \\
$211$  & $ \left \langle  f_{-}^{\mu\nu}f_{-\mu\nu}  \right \rangle $ & $\left \langle  f_{-}^{\mu\nu}f_{-\mu\nu}  \right \rangle $\\
$212$  & $i [f_{-}^{\mu\nu},{f_{-\mu}}^{\lambda}]\sigma_{\nu\lambda} $& $ 2 [f_{-}^{\mu\nu},{f_{-\mu}}^{\lambda}][S_{\nu},S_{\lambda}] $ \\
$213$  & $ \{f_{-}^{\mu\nu},{f_{-\mu}}^{\lambda}\}D_{\nu\lambda}+h.c.$ & $-4m^{2}\{f_{-}^{\mu\nu},{f_{-\mu}}^{\lambda}\}\upsilon_{\nu}\upsilon_{\lambda}$\\
$214$  & $ \left \langle  {f_{-\mu}}^{\lambda}f_{-\nu\lambda}  \right \rangle D^{\mu\nu} +h.c.$ & $-4m^{2}\upsilon^{\mu}\upsilon^{\nu} \left \langle  {f_{-\mu}}^{\lambda}f_{-\nu\lambda}  \right \rangle$\\
$215$  & $i [f_{-}^{\mu\nu},f_{-}^{\lambda\rho}]\sigma_{\mu\lambda}D_{\nu\rho} +h.c.$ & $-8m^{2}[f_{-}^{\mu\nu},f_{-}^{\lambda\rho}][S_{\mu},S_{\lambda}]\upsilon_{\nu}\upsilon_{\rho}  $\\
$216$  & $\left \langle \chi \chi^{\dagger} \right \rangle $ & $\left \langle \chi \chi^{\dagger} \right \rangle $ \\
$217$  & $\left \langle F_{R}^{\mu\nu}F_{R\mu\nu} +F_{L}^{\mu\nu}F_{L\mu\nu}\right \rangle  $ & $ \left \langle F_{R}^{\mu\nu}F_{R\mu\nu}  + F_{L}^{\mu\nu}F_{L\mu\nu} \right \rangle $\\
$218$  & $\left \langle {F_{R\mu}}^{\lambda}F_{R\nu\lambda} +{F_{L\mu}}^{\lambda}F_{L\nu\lambda} \right \rangle D^{\mu\nu} +h.c. $ & $-4m^{2} \upsilon^{\mu}\upsilon^{\nu}\left \langle {F_{R\mu}}^{\lambda}F_{R\nu\lambda}+ {F_{L\mu}}^{\lambda}F_{L\nu\lambda} \right \rangle $ \\
\end{longtable}
\bibliography{references}
\end{document}